\documentclass{sig-alternate}%,twocolumn,10pt]{article}
\usepackage{usenix,lastpage,epsfig,cite,endnotes,framed,url,caption,subcaption,graphicx}
\usepackage{amsmath,xcolor,soul,algpseudocode,lastpage,algorithm,xspace,tabularx}
\PassOptionsToPackage{hyphens}{url}
\usepackage{cleveref}

\expandafter\def\expandafter\UrlBreaks\expandafter{\UrlBreaks
  \do\a\do\b\do\c\do\d\do\e\do\f\do\g\do\h\do\i\do\j%
  \do\k\do\l\do\m\do\n\do\o\do\p\do\q\do\r\do\s\do\t%
  \do\u\do\v\do\w\do\x\do\y\do\z\do\A\do\B\do\C\do\D%
  \do\E\do\F\do\G\do\H\do\I\do\J\do\K\do\L\do\M\do\N%
  \do\O\do\P\do\Q\do\R\do\S\do\T\do\U\do\V\do\W\do\X%
  \do\Y\do\Z}

\newcount\Comments  \Comments=1   % TODO: set to 0 for final version
\newcommand{\kibitz}[2]{\ifnum\Comments=1\textcolor{#1}{#2}\fi}

\providecommand{\vs}{vs. }
\providecommand{\ie}{\emph{i.e.,} }
\providecommand{\eg}{\emph{e.g.,} }

\providecommand{\etal}{\emph{et al.\xspace} }   %Removed trailing space here; usually want non-breaking space with following reference
\providecommand{\myparab}[1]{\smallskip\noindent\textbf{#1} }

\providecommand{\systemname}{Castle\xspace}
\providecommand{\rts}{real-time strategy\xspace}

\newenvironment{packeditemize}{\begin{list}{$\bullet$}{\setlength{\itemsep}{0.2pt}\addtolength{\labelwidth}{-4pt}\setlength{\leftmargin}{\labelwidth}\setlength{\listparindent}{\parindent}\setlength{\parsep}{1pt}\setlength{\topsep}{0pt}}}{\end{list}}

\newcommand{\squishenum}{
   \begin{enumerate}{}
    { \setlength{\itemsep}{0pt}      \setlength{\parsep}{0pt}
      \setlength{\topsep}{3pt}       \setlength{\partopsep}{0pt}
      \setlength{\leftmargin}{1.5em} \setlength{\labelwidth}{1em}
      \setlength{\labelsep}{0.5em} } }

\newcommand{\squishlist}{
   \begin{list}{$\bullet$}
    { \setlength{\itemsep}{0pt}      \setlength{\parsep}{3pt}
      \setlength{\topsep}{3pt}       \setlength{\partopsep}{0pt}
      \setlength{\leftmargin}{1.5em} \setlength{\labelwidth}{1em}
      \setlength{\labelsep}{0.5em} } }

\newcommand{\squishlisttwo}{
   \begin{list}{$\bullet$}
    { \setlength{\itemsep}{0pt}    \setlength{\parsep}{0pt}
      \setlength{\topsep}{0pt}     \setlength{\partopsep}{0pt}
      \setlength{\leftmargin}{2em} \setlength{\labelwidth}{1.5em}
      \setlength{\labelsep}{0.5em} } }

\newcommand{\squishend}{
    \end{list}  }

\newcommand{\squishenumend}{
	\end{enumerate}	}

\begin{document}
%don't want date printed
\date{}

\title{{\Large \bf Games Without Frontiers: Investigating Video Games as a Covert Channel}}
\author{
%[ Paper \# 253 -- 12 pages]
{\rm Bridger Hahn}\\
Stony Brook University
\and
{\rm Rishab Nithyanand}\\
Stony Brook University
\and
{\rm Phillipa Gill}\\
Stony Brook University
\and
{\rm Rob Johnson}\\
Stony Brook University
}

\maketitle

% Use the following at camera-ready time to suppress page numbers.
% Comment it out when you first submit the paper for review.
%\thispagestyle{empty}

\subsection*{Abstract}
The Internet has become a critical communication infrastructure for citizens to
organize protests and express dissatisfaction with their governments.  This fact
has not gone unnoticed, with governments clamping down on this medium via
censorship, and circumvention researchers working to stay one step
ahead.  

In this paper, we explore a promising new avenue for covert channels:
\rts video games. Video games have two key features that make them
attractive cover protocols for censorship circumvention.  First, due
to the popularity of gaming platforms such as Steam, there are a lot
of different video games, each with their own protocols and server
infrastructure.  Users of video-game-based censorship-circumvention
tools can therefore diversify across many games, making it difficult
for the censor to respond by simply blocking a single cover protocol.
Second, games in the same genre have many common features and
concepts.  As a result, the same covert channel framework can be
easily adapted to work with many different games.  This means that
circumvention tool developers can stay ahead of the censor by creating
a diverse set of tools and by quickly adapting to blockades created by
the censor.

%% We show how these facts can be leveraged to design a coding scheme for 
%% translating data into game commands in a way that is general across 
%% games and requires little per-game customizations. 
%\textcolor{red}{This is a very bold assertion that could rankle a reviewer.
%  We need to rephrase it in a way that make it impossible to argue
%  with.}
%As a result, creating new video-game-based 
%covert channels can be made significantly easier than detecting them, making this work 
%the first proposal that provides an asymmetry in favor of developers.

We demonstrate the feasibility of this approach by implementing our
coding scheme over two \rts games (including a very popular
closed-source game). We evaluate the security of our system prototype
-- \systemname -- by quantifying its resilience to a censor-adversary,
its similarity to real game traffic, and its ability to avoid common
pitfalls in covert channel design.  We use our prototype to
demonstrate that our approach can provide throughput which is amenable
to transfer of textual data, such at e-mail, SMS messages, and tweets,
which are commonly used to organize political actions.

%\textcolor{red}{I want to tie in what we accomplished with the goals
%  we set out at the beginning, but maybe the last clause is too much?}

% translates data into game moves, that is general across games 
%and keep per-game customizations low. We evaluate these channels in
%design and prototype covert channel system
%using real-time strategy video games as a cover protocol.  Our system
%encodes data as a series of in-game moves and leverages off-the-shelf
%desktop automation to drive a game client without requiring access to
%source code.  We demonstrate the extensibility of this approach by
%prototyping on both an open source game, and the popular closed source
%game, Age of Empires.  We evaluate the security of our system based on
%its ability to match architectural properties, reliability tolerance,
%as well as packet and flow properties of regular game traffic.
%%can add something more concrete here...
%Our analysis and experiments highlight that video games significantly
%raise the bar in terms of analysis an ISP would need to do to
%distinguish regular game-play from the covert channel. In terms of
%performance, our prototype delivers adequate throughput for delivering
%text-based messages and can support higher data rates via a tradeoff
%with detectability.

%
% Metapoint: are these video games? online computer games? let's pick a term
% and stick with it.
%

\section{Introduction}\label{sec:intro}

The Internet has become a critical communication infrastructure for
citizens to obtain accurate information, organize
political actions~\cite{arab-spring-twitter}, and express dissatisfaction with
their governments~\cite{iran-twitter}.  This fact has not gone
unnoticed, with governments clamping down on this medium via
censorship~\cite{oniBurma, oniDuBlocks2008, oniID},
surveillance~\cite{nsa-prism} and even large-scale Internet take
downs~\cite{dainotti2011, iroot, pakistan}. The situation is only
getting worse, with Freedom House reporting 36 of the 65 countries
they survey experiencing decreasing levels of Internet freedom between
2013 and 2014~\cite{freedomhouse}.

%% This is not the topic sentence of this paragraph. -- rob
%%
%% Researchers working towards circumventing online information controls are
%% currently engaged in a resource-imbalanced arms race with censors. 
Researchers have responded by proposing several
\emph{look-like-something} censorship circumvention tools, which aim
to disguise covert traffic as another protocol to evade detection by
censors. This can take two forms: either mimicking the cover protocol
using an independent implementation, as in
SkypeMorph~\cite{skypemorph} and StegoTorus~\cite{stegotorus}, or
encoding data for transmission via an off-the-shelf implementation of
the cover protocol, as in FreeWave~\cite{freewave}.

This has created an arms race between censors and circumventors.  Many
censors now block Tor~\cite{tor-blocked}, so Tor has introduced support for
``pluggable transports'', i.e. plugins that embed Tor traffic in a
cover protocol.  There are currently six pluggable transports deployed
in the Tor Browser Bundle~\cite{tor-pt}.  Censors have already begun
blocking some of these transports~\cite{tor-pt-blocked}, and some censors have
gone so far as to block entire content-distribution networks that are
used by some circumvention systems~\cite{gfw-blocks-cdns}.

%% We have to state this carefully.  We don't want to make it sound like 
%% ``look-like-something'' is impossible. -- rob
%%
%%
Furthermore, recent work has shown that care must be taken when
designing and implementing a look-like-something covert channel.
For example, Houmansadr \etal showed that, when a covert channel
re-implements its cover protocol, the copy is unlikely to
be a perfect mimic of the original protocol, and a censor can use the
differences to recognize when a client is using the covert
channel~\cite{parrot-is-dead}.  However, Geddes \etal demonstrate that
even running the cover application is not enough to avoid detection by
censors~\cite{cover-your-acks} -- i.e., approaches like FreeWave may
be detected via discrepancies between the application's regular
behavior and its behavior when being used as a covert channel.  They
classify these discrepancies into three categories: (1) architectural
mismatches between communication patterns of the application when it
is acting as a covert channel \vs regular operation, (2) channel
mismatches between reliability requirements of the application and the
covert traffic and (3) content mismatches where the packet contents of
the application differ because of the covert traffic being sent in
place of regular application traffic.

%% Geddes \etal \cite{cover-your-acks}
%% argue that, even if a covert channel uses an off-the-shelf
%% implementation of the cover protocol, it should use a cover protocol
%% with similar loss tolerance, architecture, and content.  Mismatches in
%% any of these dimensions can make the covert channel easy to detect or
%% block.

In light of this state of affairs, this paper argues that video games
have several features that make them an attractive target for covert
channel development.

\begin{figure} \centering \includegraphics[trim=0cm 0cm 0cm
2cm,clip=true,width=.48\textwidth]{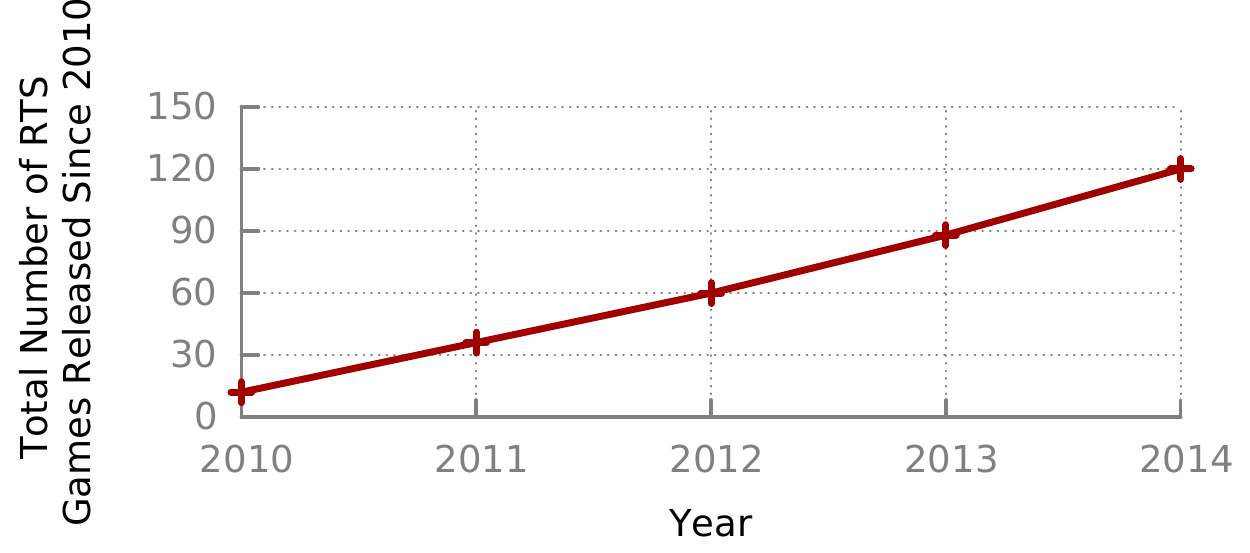}
\caption{Growth of the \rts game video-game genre from 2010 to
2014.}\label{fig:rts-steam} \end{figure}

\textbf{There are many games available, enabling circumvention tool
  developers to create a diverse set of circumvention tools.}  The
number of RTS games has grown rapidly in the last few years, as shown
in \Cref{fig:rts-steam}.  This growth has been driven in part by the
democratization of game publishing, as embodied in game distribution
platforms such as Steam~\cite{steam}.  Each game uses its own network
protocol and infrastructure, so the censor cannot simply block all
games using a single rule.  Censorship circumventors can use this
large body of games to avoid a censor's attempt to block any
particular game.

\textbf{Video games share common elements, making it possible to use a
  single framework across many games.}  For example, most Real-Time
Strategy (RTS) games have the notions of buildings, units, and rally
points, and censorship circumvention tools that encode information by
interacting with these objects can be easily ported from one RTS game
to another.  Many games also feature replay logs and similar user
interfaces, enabling covert channel frameworks that are only loosely
coupled to any particular game.

\textbf{Circumvention tools can re-use off-the-shelf game
  implementations to avoid the pitfalls identified by Houmansadr
  \etal}  Since games have features that make it relatively easy to
automate interaction with the off-the-shelf implementation,
circumvention tool developers do not need to re-implement the game (or
its network protocol), ensuring that the censorship tool uses a
faithful implementation of the game protocol.

\textbf{Games provide good architectural, channel, and content matches to
  censorship circumvention tools for textual communications.}  Many
game support both peer-to-peer and server-based gaming sessions, so
they can adapt to whichever is better for the circumvention tool.
Games must maintain synchronized state, so they are loss sensitive,
avoiding the pitfalls that Geddes \etal found in
FreeWave~\cite{freewave}.  Finally, games send frequent small packets,
matching textual communications.

\textbf{Games have built-in security features that can support secure
  covert channels.}  For example, most games include encryption and
authentication to prevent cheating. Many games also support
password-protected sessions, which can prevent application-level
attacks in which the censor attempts to identify covert channels by
joining the game.

%% \begin{itemize}
%%   \item There are numerous games, which can be used to develop a large
%%     number of diverse covert channels, making blocking all of them
%%     difficult.
%%   \item Many video games share common elements, so it may be possible
%%     to develop a single framework that can be used with many games,
%%     reducing development costs for circumvention tool authors.
%%   \item Covert channels can encode messages as in-game actions,
%%     re-using off-the-shelf game implementations to avoid the pitfalls
%%     identified by Houmansadr \etal.
%%   \item Games require reliable delivery, support both server-based and
%%     peer-to-peer architectures, and have other features identified by
%%     Geddes \etal that make them a good match for covert channel
%%     development.
%%   \item Games have built-in security features, such as encryption and
%%     password-protected sessions, that are useful for covert channel
%%     development.
%% \end{itemize}

By lowering the development cost of creating new covert channels,
video games may create an asymmetry that censorship circumventors can
use to win the arms race against censors.  Censors can respond to
look-like-something circumvention tools by blocking the cover protocol
entirely or attempting to distinguish legitimate uses of the protocol
from uses by the covert channel.  If developing such blocking tools is
time consuming, but the circumvention tool developers can quickly move
their tool to a new cover protocol, then there will almost always be 
effective circumvention tools available for end users.

%% With these prior results in mind, we consider a new avenue for covert channels:
%% \rts video games. Similar to Skype, video games provide another opportunity to 
%% encode data using off-the-shelf game implementations.
%% However, the current video game landscape also presents three advantages that
%% distinguish games from other covert channels and make them  amenable to winning
%% the arms race between censors and circumvention tool developers. 

However, we must answer several questions to understand the
feasability of using video games for covert channels:
\begin{itemize}
  \item Can video games support good covert channel bandwidth?
  \item Can we encode data in the video game so that the censor cannot
    distinguish regular game play from covert channel sessions?
  \item Can we develop a covert channel framework that can be quickly
    adapted to new games?
\end{itemize}

To answer these questions, we have built \systemname, a prototype
video-game-based covert-channel framework.  \systemname encodes data
as player actions in an RTS game.  \systemname uses desktop-automation
software to execute these actions in the game.  The video game
software transmits these moves to the other players in the same gaming
session, who then decode the message and send replies in the same way.

\systemname can be easily adapted to new RTS games.  Our current
prototype supports two such games: ``0-A.D.''  \cite{0ad} and an
extremely popular (over three million copies sold) closed-source game
that we will refer to as ``Aeons''.  It took a single undergrad less
than six hours to port \systemname from 0-A.D. to Aeons.  

\systemname is easy to port to new RTS games for two reasons.  First,
\systemname uses only features, such as buildings, units, and ``rally
points'', that are nearly universal to RTS Games.  Thus the high-level
architecture and encoding scheme can be re-used across games.  Second,
\systemname is only loosely coupled to game internals.  For example,
\systemname uses desktop-automation software to execute game actions
through the game's standard graphical user interface.  Similarly, the
\systemname decoder reads actions from the game's replay log, which
means that it does not need to understand the game's network protocol
or other internals.  For many games, development is made evern easier
by the ready availability of code to parse their replay logs.

\systemname offers good bandwidth for text-based communications.  Our
current prototype provides between 50 and 200 B/s of bandwidth,
depending on configuration parameters.  \systemname has about
100x more bandwidth than other proposed game-based covert channels~\cite{Zander-1,Zander-2,rook}\footnote{ Despite the similarity of their names and their
  common use of video games, Rook and \systemname were developed
  independently and have quite different goals.  See
  \Cref{sec:discussion} for details.}  .
With some game-specific
tuning, the Aeons version can deliver over 400 B/s.  Even 50 B/s is
sufficient for email, SMS messages, and tweets, which are widely used
organizational tools among political activists.  There are also
several ways to potentially increase \systemname's bandwidth (see
\Cref{sec:discussion} for details).

\systemname's design makes it resilient to most classes of attacks.
Since \systemname uses the underlying game to transmit data, an
attacker cannot use simple IP- or port-based blocking to block
\systemname without blocking the game entirely.  When used with games
that encrypt and authenticate their traffic, an attacker cannot use
deep packet inspection to distinguish \systemname traffic from regular
game traffic.  Encryption and authentication also preclude simple
packet injection or manipulation attacks.  Since games use network
communication to synchronize their state, they are loss sensitive,
unlike some VoIP protocols.  Thus \systemname cannot be distinguished
from regular gaming sessions through selective packet delay or
dropping attacks.  Finally, when used with password-protected gaming
sessions, \systemname is immune to application-level attacks, such as
the censor attempting to join the same gaming session to observe the
player's in-game actions.

We evaluate \systemname's security against statistical
traffic-analysis attacks by applying several previously published
classifiers.  We find that packet sizes and interpacket times of
\systemname's traffic deviate from those of regular human-driven game
play by the same amount that different human player's traffic differ
from eachother.  We also find that the Liberatore~\cite{liberatore},
Hermann~\cite{Herrmann}, and Shmatikov~\cite{Shmatikov} classifiers cannot
distinguish \systemname traffic from regular game traffic with success
much better than random guessing.

Together, these results show that video games offer promise as a
target for covert channel development and that video games may enable
circumention tool developers to gain the upper hand in the arms race
against censors.

%I generally wouldn't subsection this.
\myparab{Paper outline. }
%% In Section \ref{subsec:arms-race}, we provide a brief
%% overview of the current state-of-the-art in \emph{look-like-something} 
%% censorship circumvention systems and explain how \systemname differs from
%% other video game based covert channels. 
In Section \ref{sec:threat}, we present
the adversary model that we consider in this paper. 
Section \ref{sec:system} provides some background on \rts games, details the
properties that makes them favorable for use as cover protocols in covert 
channels, and explains how \systemname makes use of each of these for sending
and receiving covert data. In Section \ref{sec:implementation}, we provide 
details on our publicly available implementation of \systemname. 
Following this, we describe our evaluation framework in Section \ref{sec:setup}.
In Sections \ref{sec:security} and \ref{sec:performance}, we present the results 
of \systemname's security and performance evaluation, respectively.
In Section \ref{sec:discussion}, we discuss the impact that \systemname makes 
on the currently on-going censor-developer arms race, modifications that may be
made to \systemname for additional throughput gains, and compare the primary
design principles of \systemname with it's most similar counter-parts. Finally,
in Section \ref{sec:conclusions}, we draw our conclusions and provide a link to
a video demonstration of \systemname.

\section{Adversary and Threat Model}\label{sec:threat}
In this paper, we consider a network-level censor (e.g., an ISP) who is able 
to perform analysis over all traffic that it forwards from or to clients 
within its network. It may also perform manipulations (e.g., dropping packets, 
injecting packets) of the network traffic via on-path or in-path middleboxes. 
The adversary may also take an active approach and probe application endpoints 
or otherwise interact with the game. 

In this section, we overview the capabilities of the censor that we aim to 
evade using \systemname. We describe the resilience of \systemname to these 
adversary behaviors in Section~\ref{sec:security}.

%perform probing of the application. 
%Unlike most previous work, we consider censors that may perform 
%proactive blocking, and active and passive analysis to detect the identities 
%of, or block clients using censorship circumvention tools, while not degrading 
%the availability of services to other clients.

\subsection{Network traffic attacks}

%\myparab{Passive analysis capabilities. }
\myparab{Passive analysis.} We consider censors that are able to perform stateless and stateful passive 
analysis of traffic at line rate. In particular, the censor is able to 
perform the following passive analyses to detect the use of a circumvention
tool:
\begin{packeditemize}
\item \textbf{IP and port filtering: } The censor can observe the IP addresses and port numbers 
of connections on their network (\eg using standard tools like Netflow~\cite{netflow}). 
%If a video game uses 
%specific servers (IP addresses) or port numbers, the ISP is easily able to identify game flows.

%this description is active behavior (blocking)
%The censor can identify and block all 
%client interactions to specific IP addresses and ports.
\item \textbf{Deep-packet inspection: }The censor may look for specific 
patterns in packet headers and payloads (\eg application payloads indicative of a specific game).
\item \textbf{Flow-level analysis: }The censor may perform statistical
analyses of flow-level characteristics such as inter-packet times and packet 
sizes).
\end{packeditemize}

The first two of these capabilities mean that the ISP can easily detect flows related to the video game in general. For example, if the game uses a specific set of servers (IPs) or ports these flows may be easily identified. Similarly, game-specific application payloads can reveal game traffic to the ISP. The last property can reveal information about game behavior to the ISP (\eg rate of play). A circumvention system must avoid perturbing these features to remain undetected and unblocked. 

\myparab{Active manipulations.}  
In order to detect and/or disrupt the use of censorship circumvention tools, censors may 
perform a variety of active manipulations on suspicious connections that transit  
its network. 
In particular, the censor may drop, insert, or delay packets. 
Additionally, they may also modify the packet contents and headers.
The adversary may perform these manipulations to observe the behavior of flow endpoints to distinguish legitimate game traffic from the covert channel. They may also use these actions to block covert connections (\eg sending TCP RST packets, or dropping traffic).

\subsection{Application layer attacks}

%\myparab{Proactive blocking capabilities. }
In the context of detecting \emph{look-like-something} covert channels, 
censors may take additional actions outside the scope of standard 
active and passive analysis. Specifically, they may interact with the application 
that the covert channel aims to hide within. They may join game servers and observe 
games in progress (\ie who is playing with whom). They may observe properties of the games (\eg map state, player move behaviors) or join 
and interact with game players if the game is not password protected. 

%Specifically, they may create custom tools 
%(e.g., they may create tools that use the covert channel or cover protocol 
%to identify other users of the covert channel). 

%this can also fall under the active manipulations in network layer. blocking is a network layer function ... or at least best fits in that bucket
%Additionally, in some cases
%they may block all uses of the cover protocol itself -- legitimate or 
%otherwise.

\myparab{Censor limitations. } We impose some limitations on the
computational capabilities of censors. While they have a large amount
of computational resources, they are still unable to decrypt encrypted
communication channels and guess high entropy passwords.  We also
assume that the censor does not have a back door into the game or game
servers.  For example, we assume the censor is not able to break into
the game servers (e.g. by exploiting a buffer overflow or other bug).
We also assume that the operators of the game servers do not cooperate with the censor,
e.g. they do not allow the censor to see other user's private game
state.

\section{The \systemname  Circumvention Scheme}\label{sec:system}
%\FIXME{\st{Text overviewing the architecture (note: this section is more high level 
%architecture/idea, implementation details are later). A figure showing the 
%structure of the system with the client in censoring country with game client, 
%talking to game client in free country Game client in free country is connected 
%to Web proxy. Game client in censoring country is connected to a browser plug 
%in.}}

\systemname aims to demonstrate that secure and low-bandwidth
\emph{look-like-something} defenses are possible via interactive channels such
as real-time strategy video-games. In this section, we provide a background on
real-time strategy games and highlight key properties of these games that enable
\systemname to create covert channels that generalize to the entire
genre. We then describe how \systemname encodes, sends, and receives data.

\subsection{Real-time strategy games}\label{subsec:rts}

%We now present a brief background on real-time strategy games.
Real-time strategy games are a genre of video games that center
around the idea of empire-building. Typically, the goal is for a player to
assert control over enemy territory through a combination of military conquest
and economic maneuvering. Below we highlight commands and features that are 
common to a large majority of \rts games (Table \ref{tab:featuresgames}).

\begin{packeditemize}
\item \textbf{Units: }Real-time strategy games allow players to \emph{create} and
\emph{train} a large number of units (\eg human characters, livestock,
machinery). Units may perform many actions \eg in 17 of the Top 20 best-selling
\rts games, a unit can be instructed to move to a location on the map by left-
clicking it and then right clicking the destination location on the map.

\item \textbf{Buildings: }Players may construct a number of buildings over the
course of a game. Buildings are required to train certain units and research new 
technologies. For instance, barracks may be required to train infantry.
Some buildings produce new units of other types.
In most \rts games, unit-producing buildings can be assigned a rally point -- 
i.e., a location on the map at which all units created by the building will 
assemble. This command is available in 17 of the Top 20 best-selling \rts
games.

\item \textbf{Maps and map editors: }Real-time strategy games are set in a 
landscape covered by plains, forests, mountains, and/or oceans. Most real-time 
strategy games allow users to create their own maps and modify existing maps for use 
within the game. Map editors released by the publisher or the modding community
are available for 17 of the Top 20 \rts games.

\item \textbf{Replay files: }Players may be given the option to record all moves
and commands issued by themselves and other players in the game. This is used to
replay or watch previously played video-games. When this option is enabled, the 
game writes, in real-time, all commands issued in the game to a replay log that 
may be in a proprietary format. Replay file decoders are available for 11 of 
the Top 20 \rts games.
\end{packeditemize}

In addition to the above elements, many commercial \rts games also possess
the following networking and security properties that are advantageous for use as cover
protocols for covert channels.

\myparab{Network communications.} For scalability reasons, real-time strategy
games do not broadcast state information to all players in the game. Instead, 
they pass commands issued by the players in fixed intervals (\eg 100 ms). These 
commands are then simultaneously simulated in each game client.  This allows 
clients to execute the game identically, while requiring little bandwidth 
\cite{aoe-paper}. As a consequence, any data encoded as an in-game command is 
received as such, by players at the other end.

Additionally, while most \rts games make use of UDP channels for command 
communication, reliability is implemented in the application layer. This makes
many active traffic manipulation attacks described in previous work 
\cite{cover-your-acks} ineffective.

In terms of network architecture,  real-time strategy games may take two forms,
with players joining a common game hosted on a game server (\eg servers hosted
by game publishers such as Microsoft, Blizzard, Electronic Arts, etc.), or 
connecting directly to each other in a peer-to-peer mode. 
Therefore, any covert channel system utilizing video games as a cover, can employ 
whichever is the dominant mode of operation and shift from one to the other if 
required, to evade censorship.

\myparab{Security considerations.} %As mentioned in Section \ref{sec:intro}, 
Real-time strategy games often implement several security mechanisms in order to prevent 
cheating in multi-player game sessions. These include encrypting and authenticating
the communication channel that carries player commands, verifying the 
consistency of the game state with other clients in the game, and restricting access to game sessions via a password.

These security mechanisms have several vital consequences for their use as 
covert channels. First, since the game command channel is encrypted, passive 
adversaries are unable to view commands issued by players in a game by simply 
observing network traffic. Second, the presence of authenticated channels and
game-state verification algorithms prevents active attackers from using
falsified game packets to interact with, or observe other clients on the game 
servers.

\myparab{Commonalities between \rts games.} Our prototype, \systemname, leverages the common command structure, map 
design capabilities, and tools for decoding saved games and replays generated 
by real-time strategy games. A survey of real-time strategy games reveals that 
11 of the top 20 best-selling games of all-time also include these features 
(Table \ref{tab:featuresgames}).

%% This doesn't go here. -- rob

%% To demonstrate the ease at which covert channels can be deployed over games
%% within a genre, once the our prototype -- \systemname -- was developed to
%% implement a covert channel over 0 A.D., it took a bright undergrad less than
%% six hours to re-configure and implement it over a very popular closed-source
%% \rts game having the features listed in Table \ref{tab:featuresgames}.

%% Indeed, the data encoding scheme and input mechanism are virtually identical
%% between the games with the majority of effort spent determining how to decode
%% replay files on the receiving client.

\begin{table}[ht]
  \centering
  \begin{tabular}{|p{4cm} |c|}
  \hline
  Feature & Number of Games\\\hline
  Common Comands (\texttt{MOVE-UNITS} or \texttt{SET-RALLY-POINT}) & 17 of 20\\
  \hline
  Map Editors & 17 of 20\\
  \hline
  Replay Decoding Tools & 11 of 20\\
  \hline
  \end{tabular}
  \caption{Real-time strategy %we're burning a line for the cite anyways ... 
  game features used by \systemname and the number of games in the
Top 20 best-sellers of all-time that possess them. \cite{pc-games-wiki} }
  \label{tab:featuresgames}
\end{table}

\subsection{Building game-based covert channels}
%\systemname exploits two key properties of real-time strategy
%games to create a covert channel mechanism that is general to the majority of
%games in this category:

%%%%% I don't think we should even discuss this.  -- rob

\myparab{Straw-man approach. }One may consider establishing covert communication
channels via the in-game voice and text chat channels. However, this approach
has several drawbacks. First, previous work shows that encoded data is easily 
distinguishable from human audio communication~\cite{cover-your-acks, parrot-is-dead}.
%has shown that encoding data to immitate human audio communication is very
%hard to do. 
Furthermore, voice communication channels are fairly uncommon in the
\rts game genre. Second, while game data is encrypted, it is often the case that text communication channels 
are left unencrypted. Finally, while one may expect a fairly constant stream
of human issued in-game commands in a \rts game, it is rare to have long text or audio 
communication while playing the game. These factors allow covert channels built
on these approaches to be either difficult to implement and extend, or to be trivially 
detected by an adversary, or both.

\myparab{The \systemname approach.} In order to create a covert channel mechanism that 
is general to the majority of games in the \rts genre, \systemname exploits two key 
properties.

\begin{packeditemize}

\item Most real-time strategy games share a common set of actions. Specifically,
the ability to select buildings and assign a location where units
created/trained in a building should go. This location is called a ``rally
point,'' and we denote the command of setting the rally point for units created in a given
building by {\tt SET-RALLY-POINT}. Games also provide the ability to move a
selected unit to a given location (denoted by the {\tt MOVE} command). Thus, any
encoding that translates data into a combination of unit/building selections and
these primitives will be general across most games in this class.

\item Most real-time strategy games provide a replay option which saves every
players' moves to disk (for later playback). Therefore, all in-game commands are
written to disk where they can easily be read and decoded in real-time.

\end{packeditemize}

\begin{figure}[t]
    \centering
    \includegraphics[width=0.48\textwidth,trim=1cm 4cm 1cm 4cm, clip=true]{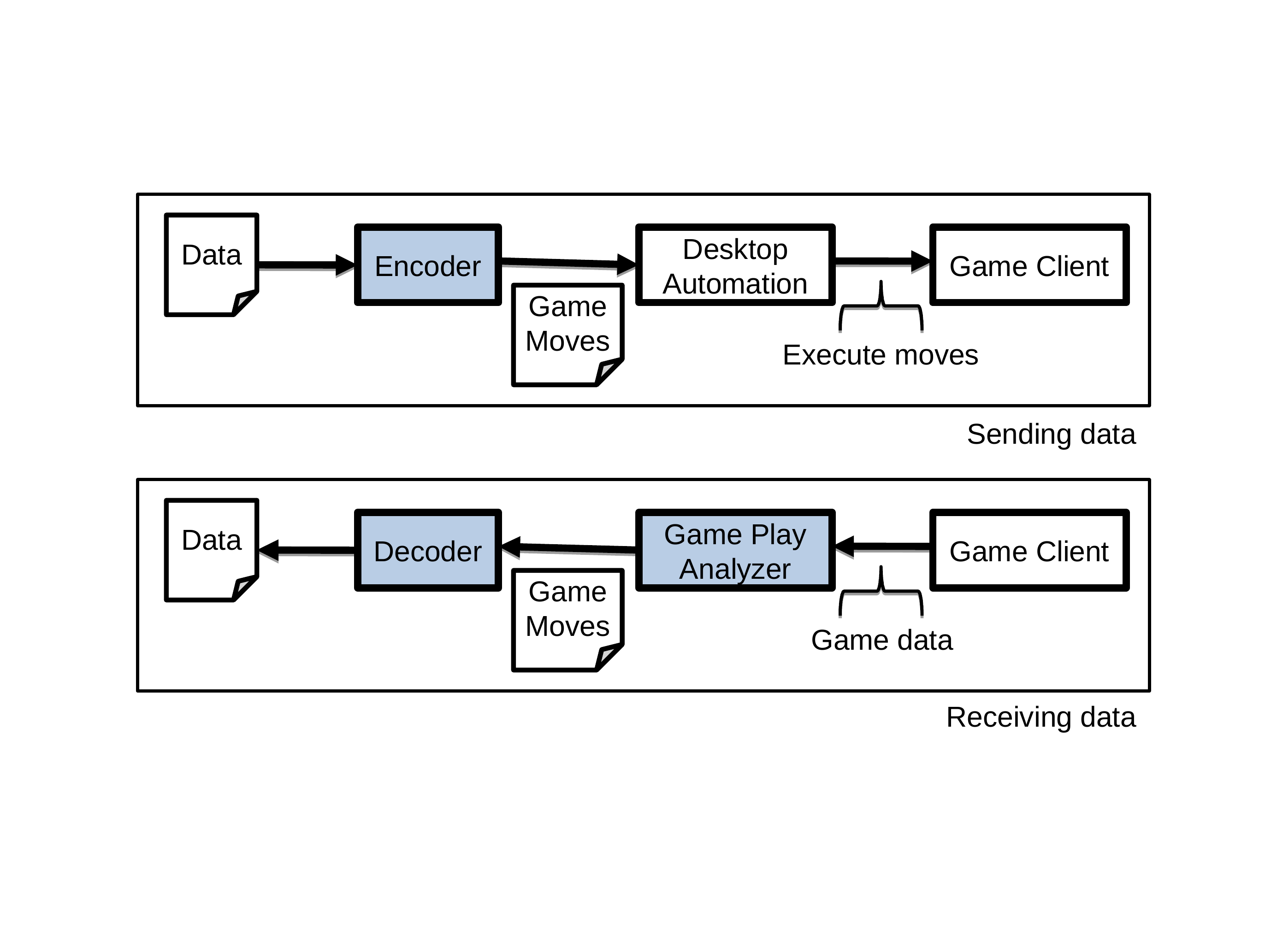}
    \caption{Overview of data flow for sending an receiving in \systemname.
Shaded components are implemented as part of \systemname while the others use
existing off-the-shelf software.}
    \label{fig:overview-1}
\end{figure}

\systemname consists of two main components to send and receive
 data. These are illustrated in Figure~\ref{fig:overview-1}. Sending is done 
 by encoding data into game commands and then executing them within the 
 game using desktop automation. The receiving process monitors the log 
 of game commands and decodes this list to retrieve data sent via the system.

\begin{figure}[t]    
\centering
    \includegraphics[width=0.48\textwidth,trim=0cm 5cm 1cm 5cm, clip=true]{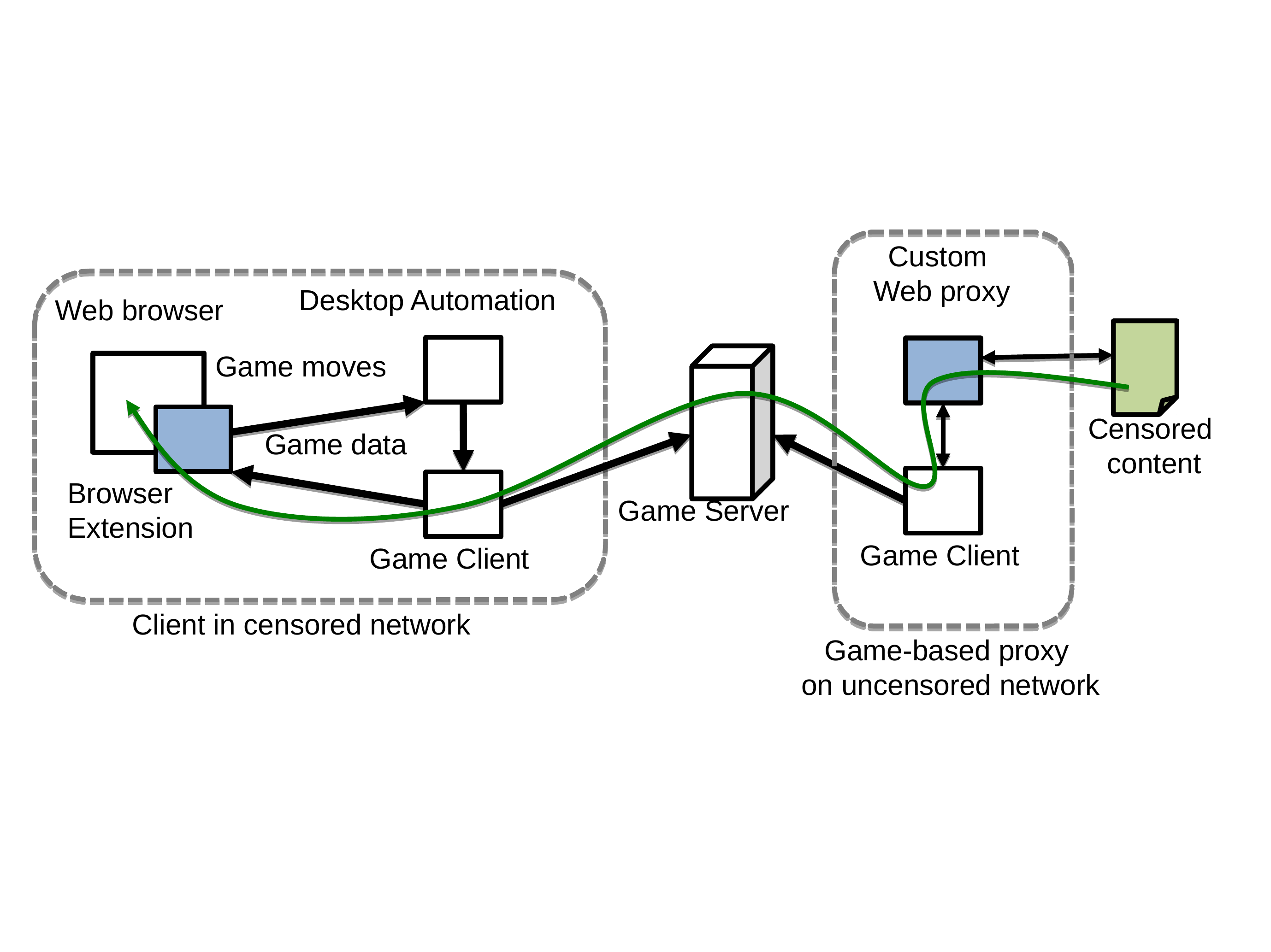}
    \caption{Overview of how \systemname can be used as a proxy  for clients
within censoring countries. A game client outside of the censoring region acts
as a proxy between the game and the censored content (\eg Web pages).}
    \label{fig:overview-2}
\end{figure}

Figure~\ref{fig:overview-2} overviews how the \systemname system could be used
to relay data from outside of a censored region to a client within the region.
The client first installs \systemname (\eg as a browser extension).  The
\systemname client then initiates a game through a game lobby (or directly with
the client outside of the censoring region). The client in the censoring region
can then encode and send data (\eg Web requests) as game moves that can be
decoded by the client outside of the censoring region. The game client outside
of the censoring region can then act as a proxy to retrieve censored content and
send it via \systemname to the client in the censoring region.

%\subsection{The games}
%\rish{I think I'm going to move this subsection to the implementation details
%section.}
%\FIXME{describe the two video games. What are they, what are the basic actions/goals. 
%What is the architecture of the game system/servers.}

\subsection{Encoding data into game commands}\label{subsec:encoding}
%\FIXME{Add a figure to illustrate the encoding process}

\systemname relies on the ability of the player to select units and buildings
and set rally points to encode data. A naive encoding may consider selecting
each unit and directing it to a different point on the game map to encode a few
bytes of information per unit. However, in  preliminary experiments, we observed
that this approach resulted in a covert channel that could not match the
properties of the original game traffic (moving O(100s) of units to distinct
locations is not a usual action for players). 
%Thus, we needed to find another
%way to encode data in \systemname. 

Encoding in \systemname is accomplished, without inflating the amount of game
data transferred, using the following scheme. First, the participants in
\systemname download a custom map (distributed via game forums or stores such 
as Steam) which contains either $n$ immobilized units (e.g., units placed in
unit sized islands, within walls, etc.) or $n$ unit producing buildings (e.g.,
barracks, stable, etc.). The \systemname sending process then encodes data by
selecting a subset of these $n$ units and executing either a {\tt MOVE} command
in the case of units or {\tt SET-RALLY-POINT} in the case of buildings. While we 
discuss the encoding in the context of units and the {\tt MOVE} command, 
\systemname is easily implemented using either primitive.

Instead of using selection of each unit to represent a single bit sequence, which would
result in $\log_2(n)$ bits of data transferred per command, we use a combinatorial scheme
where we select $k$ of the $n$ units, to increase efficiency. Intuitively, the
selection of $k$ of $n$ units results in $\binom{n}{k}$ different values or
$\log_2\binom{n}{k}$ bits that may be transferred per command. We use
combinatorial number systems~\cite{knuth-comb} to convert $\log_2\binom{n}{k}$ bits of
data into a selection of $k$ of the $n$ units on the game screen. In preliminary
experiments, we found that the selection of a constant number of units per
command
resulted in traffic which was more uniform than regular game traffic. As a
result, we adjusted our scheme to select between $0$ and $k$ units for encoding
to increase variability of packet sizes. Section~\ref{sec:security} provides a more in-depth 
view of how we evaluate our similarity to actual game traffic. 

In addition to selecting the set of units, we can also select a location for all
$k$ selected units to move to. Note that since we select a single location for $k$ units
(instead of $k$ distinct locations) this does not impact the data transfer
size. Given a game map with  $m=x_{max}\times y_{max}$ potential locations we
can additionally encode $\log_2m$ additional bits of data in a given turn. 

Assuming a map with $n$ units/buildings, a maximum of $m = x_{max} \times y_{max}$ 
map locations, and a game which allows for a maximum of $k$ units/buildings to be
selected simultaneously, the game-independent encoding of covert data into a
\texttt{MOVE} or \texttt{SET-RALLY-POINT} command is done as shown in
Algorithm \ref{alg:encode}. 

\begin{algorithm}
\caption{Algorithm for encoding covert data into game commands}
\label{alg:encode}
\begin{algorithmic}
        \Function{encode}{$data$, $k$, $n$, $m$, $x_{max}$}
                \State $r \xleftarrow[]{\$} \{1, \dots, k\}$
                \State $z_1 \gets$ \Call{read}{$data$, $\log_2\binom{n}{r}$}
                \For{$i = n \to 0$}
                        \If{$\binom{i}{r} \leq z_1$}
                                \State $z_1 \gets z_1 - \binom{i}{r--}$
                                \State $selected \gets selected||i$
                        \EndIf
                \EndFor
                \State $z_2 \gets$ \Call{read}{$data$, $\log_2 m$}
                \State \texttt{(x, y)} $\gets$ ($z_2 \mod x_{max}, \lfloor
z_2/x_{max} \rfloor$)
                \State \Return $\{selected, \texttt{(x, y)}\}$
        \EndFunction

        \Function{read}{$file$, $x$}
                \State \Return next $x$ bits from $file$ in base 10.
        \EndFunction
\end{algorithmic}
\end{algorithm}

The combination of selecting between $0$ and $k$ units and setting the location 
to move to, results in an average of 

$\left(\frac{\sum_{i=1}^{k}{\log_2 \binom{n}{i}}}{k} + \log_2 m\right)$ 
bits transferred per command. 

As mentioned earlier, one may achieve higher data-rates by always selecting $k$
units, however, this causes identically sized commands and thus affects the 
packet size distribution.

%%%%% SNIP %%%%%%%%%%%%%%%%%%%%%%

\subsection{Sending covert data}\label{subsec:sending} 
%\FIXME{This is the part about how you make the players execute moves that are given by 
%your code. discuss naive option of modifying game code (issues: hard, plus games 
%are not all open source). Calibrating the input device. talk about solution using desktop 
%automation. Advantage, easily extensible to multiple games.}

Once the covert data is encoded into in-game commands, the sending process must
actually execute the commands in order to communicate them to the receiver. One way
to do this is to modify the game AI to issue commands as dictated by our encoder.
However, this is non-trivial since most games are closed-source and
viewing/modifying game code is not always an option. Even when source code is
available, the overhead of understanding the game code
and modifying the AI presents a non-trivial hurdle. Given our vision of
adaptability to the large number of available \rts games, we leverage
off-the-shelf desktop automation to execute the encoded
game commands. 
% in the game client. 
This opens the door to extending our approach to
a much larger set of games than would otherwise be possible. 

Since the map used in \systemname is custom made, the starting location of all
units is known in advance. Further, since units and buildings are immobile, 
\systemname is aware of their location at all times. The location of units on 
the game map, along with the list of commands to be executed is sufficient for 
\systemname to automatically generate a sequence of key-presses, left-clicks, 
and right-clicks to be made by the desktop automation tool. This sequence is 
then passed to the automation tool for execution.

%There are several benefits to this approach: 
%First, there is no need to reverse engineer and modify game code. This allows
%\systemname to easily integrate new real-time strategy games into its
%repertoire.
We note that, certain automation tools allow keystrokes and clicks to be sent to
windows that are not currently in focus. This ensures that \systemname does not
detract from the user experience by requiring the game window to be in
focus during data transfer periods.Finally, since automation tools allow control 
over the speed of clicks and key-presses, \systemname can be configured to either 
mimic human input speeds (lower clicks/second) or maximize throughput (higher 
clicks/second). We investigate the trade-off between these two variables in 
Sections \ref{sec:security} and \ref{sec:performance}.

\subsection{Receiving covert data}\label{subsec:receiving}
%\FIXME{How does a player watch the game board to read out data sent by the other 
%player?}
Since the receiving game client does not have the same in-game screen as the
sending client (due to each client having their camera focused on different map
locations), directly observing the commands made by the sending client via the
screen output is prohibitively complex.
Fortunately, most real-time strategy games maintain a real-time log of all
commands issued in the game to enable replaying moves or saving game state.  In
\systemname, the receiving process constantly
monitors this log file for commands issued by other participants. These commands can
then be decoded back into their original covert data via the decoding algorithm
specified in Algorithm \ref{alg:decode}.
%Appendix \ref{sec:appendix} (Algorithm \ref{alg:decode}).

\begin{algorithm}
\caption{Algorithm for obtaining covert data from game commands}
\label{alg:decode}
\begin{algorithmic}
        \Function{decode}{$selected$, \texttt{(x, y)}}
                \State $size \gets |selected|, z_1 = 0$
                \State $selected \gets \Call{sort-descending}{selected}$
                \For{$i \in selected$}
                        \State $z_1 \gets z_1 + \binom{i}{size--}$
                \EndFor
                \State $z_2 \gets (\texttt{y}\times x_{max}) + \texttt{x}$
                \State \Return $(base2(z_1)||base2(z_2))$
        \EndFunction
\end{algorithmic}
\end{algorithm}

%provide an option to view replays of entire games.
%To enable this feature, the game stores (in real-time) the history of all moves made by all game
%participants in a replay log. In \systemname, the receiving process constantly
%monitors this log file for moves made by other participants. These moves can
%then be decoded back into their original covert data via the decoding algorithm
%specified in Algorithm \ref{alg:encode-decode}.

This approach suffers from one minor drawback: replay logs for games from 
commercial studios are often stored in proprietary and undocumented formats 
that vary from game to game. However, reverse engineering the format of the 
replay logs is made significantly easier since \systemname only issues 
\texttt{MOVE} or \texttt{SET-RALLY-POINT} commands. Therefore, we only need
to understand how these commands are stored in replay logs. This can be
done by simple techniques -- e.g., sending a unit to the exact same location
multiple times allows us to obtain the byte code used to signify the
\texttt{MOVE} command, sending a unit to two locations in sequence (with each 
separated by a single pixel) allows us to obtain the bytes used to denote the
\texttt{(x, y)} destination co-ordinates, etc. Further, for many popular \rts games, 
these formats have already been reverse-engineered by the gaming/hacking community.

%\textbf{\systemname Co-ordinate Calibration: }
%The isometric perspective of the game screen poses a challenge during the
%decoding process. Specifically, the presence of a \emph{viewing angle} means
%that mapping screen co-ordinates to in-game \texttt{(x, y)} co-ordinates is not
%trivial and linear. Therefore, while the sender may have intended to move a
%unit to the screen co-ordinate \texttt{(x$_s$, y$_s$)}, the game actually logs
%the command as an order to move the unit to the game co-ordinate \texttt{(x$_g$,
%y$_g$)}. This is also the command obtained by the receiver on decoding the move
%log. In order to circumvent this problem, \systemname goes
%through a one-time (per game title) calibration process. 
%
%During this process, \systemname uses the desktop automation module to issue 
%commands to move a single unit to every pixel on the 
%game screen. Next, it uses the decoder to obtain the list of in-game commands
%issued by the automation tool. This provides a direct mapping between each 
%on-screen co-ordinate \texttt{(x$_s$, y$_s$)} and in-game co-ordinate
%\texttt{(x$_g$, y$_g$)}. This mapping is used to obtain the original location 
%bits encoded by the sender.
%
%Note that the resulting mapping from this calibration process can be shared
%across clients as long as the same display resolution and camera settings are
%maintained. 

%\subsection{Web Proxy}
%\rish{Moving to implementation section}
%\FIXME{This is connect to the game + proxies HTTP connections to the Web. Where are 
%good places to think about hosting this sort of thing? }

\section{\systemname Prototype Implementation}\label{sec:implementation}
%\rish{Let's move the calibration details to this section. Also talk about how we
%(1) construct the map, (2) use the desktop automation tool, (3) access the save
%game file, and (4) do calibration. Also talk about how all of this ties
%together. Number of lines of code / languages used (mention in the sections
%itself). Describe each of the games.}

In this section, we describe our prototype implementation of \systemname. We
prototype on two games to illustrate the extensibility of our approach.

\begin{packeditemize}
\item \textbf{0 A.D.: } An award-winning, free, open-source, 
and cross-platform \rts game made available under the GPLv2+ license, by 
Wildfire Games. 
\item \textbf{Aeons: } A best-selling (currently in the top 5 grossing
\rts games of all-time), closed-source, Windows-based \rts game.
\end{packeditemize}

Our prototype comprises of $\sim$500 LOC and was coded in a combination of 
Python and AutoHotkey (desktop automation) \cite{autohotkey} scripts. It 
includes the following components:

\myparab{Custom map: } 
To test \systemname, we created a custom game map for each of the two games. 
The map was comprised of $n$ buildings packed as tightly as possible to
facilitate our selection-based encoding. 

For 0 A.D., we created a map with $n = 1600$ buildings on a single
game screen, while for Aeons, we were only able to have $n = 435$ (owing to
larger unit sizes). For both games, a region large enough to contain 16 bits of
location data was left unoccupied. This is used to assign rally-point coordinates
to the selected buildings.

%We create a custom map for testing each game.
%For testing with \systemname, two types of maps were
%created for each game
%-- a unit based map and a building based map. 
%In the unit based map, an $x \times y$ walled grid is generated, with each cell
%within the grid containing exactly a single unit. The purpose of the walls is to 
%immobilize the units. This ensures that they cannot move even in spite of 
%being commanded to. Our unit based map contains a total of 200 walled-in units.
%In the building based map, $n$ buildings are packed as tightly together as the 
%game will allow. Our building based map contains a total of 1600 buildings.

%In both maps, a region large enough to contain 16 bits of location 
%data is left unoccupied by the units/buildings. This region is used to
%assign destination and rally-point co-ordinates to the selected units and buildings,
%respectively.

Since 0 A.D. stores maps in a simple and readable XML format, the process of map
creation was easily automated (via a Python script). This was not the case for 
Aeons which required manual generation of the map using the official GUI map 
editor. We are currently exploring automation options for map creation in 
Aeons.

\myparab{Data encoding and decoding: } Code for translating between covert 
data and in-game commands (and vice-versa) was written in under 200 lines of 
Python using the encoding and decoding described in Section 
\ref{subsec:encoding}. The output of the encoding code was a vector of 
buildings to be selected and a single \texttt{(x, y)} coordinate.

\myparab{Desktop automation: } We used the open-source desktop automation tool,
AutoHotkey, to execute the series of commands determined by the encoding scheme. 
Since the locations of all buildings and units were known, selecting and 
commanding those indicated by the encoding was straightforward. 

%We write an AutoHotkey script to execute these 
%commands, and leverage its ability to operate on windows that are not currently 
%focused (\ie in the background).

%Since the locations of all buildings is
%known, mapping the output of the move generation code to a sequence of click
%locations is trivial. For the purpose of executing the generated moves, 
%the free and open-source desktop automation tool -- AutoHotkey -- is used.

%The sequence of locations to be clicked along with the game window ID is passed
%to our AutoHotkey script. In order to issue commands to game while it is not in the
%foreground, the script issues \texttt{ControlClick} commands.

\myparab{Reading recorded game data: } We implemented code that monitored the 
log file of commands issued (maintained by the game), for both games. For 0 
A.D., this information was already made available in a simple to parse text 
file. In order to obtain this information for Aeons, the game replay file 
was parsed using tools made available by the gaming/hacking community. The 
file was then scanned to obtain each command as a vector of selected buildings 
and an \texttt{(x, y)} coordinate. The commands were then decoded to retrieve 
the originally encoded covert data. 

%Note that tools to reverse engineer game replay files are available for at least
%seven of the top ten best selling \rts games of all-time.

\myparab{Coordinate calibration: }The isometric perspective of the game screen 
posed a challenge during the decoding process. Specifically, the presence of 
a \emph{viewing angle} meant that a sender may have intended to move a
unit to the screen coordinate \texttt{(x$_s$, y$_s$)}, but the game actually 
logged the command as an order to move the unit to the game coordinate 
\texttt{(x$_g$, y$_g$)}, making this the command obtained by the receiver on 
decoding the move log. In order to avoid this problem, \systemname goes
through a one-time calibration process of mapping on-screen coordinates to 
coordinates as interpreted in the game. Note that the results of this 
calibration process can be shared across game clients that utilize the same 
in-game resolution.

%\iffalse
%During this process, \systemname uses the desktop automation module to issue
%commands to move a single unit to every pixel on the
%game screen. Next, it uses the decoder to obtain the list of in-game commands
%issued by the automation tool. This provides a direct mapping between each
%on-screen co-ordinate \texttt{(x$_s$, y$_s$)} and in-game co-ordinate
%\texttt{(x$_g$, y$_g$)}. This mapping is used to obtain the original location
%bits encoded by the sender.
%
%Note that the resulting mapping from this calibration process can be shared
%across clients as long as the same display resolution and camera settings are
%maintained.
%\fi

\section{Evaluation Setup}\label{sec:setup}
We evaluate \systemname along two axes. First, in Section \ref{sec:security}
we consider security of the \systemname by quantifying its resilience to the 
censor-adversary described in Section \ref{sec:threat} and its ability to 
avoid the mismatches highlighted by Geddes \etal~\cite{cover-your-acks}. We 
then study throughput of \systemname using the encoding scheme as laid out in 
Section~\ref{sec:system}. We also consider the effect of minor game-specific 
improvements to \systemname's throughput.

For the evaluation in Sections \ref{sec:security} and \ref{sec:performance}, 
we use our implementation of \systemname with a building-based map, using 
\texttt{SET-RALLY-POINT} commands.The evaluation was performed on Windows 8.1 
running AutoHotkey \cite{autohotkey} for automation. The game was set up in 
direct connect mode -- i.e., the two players were connected directly to each 
other via their IP address (rather than through the game lobby). Since both 
players were on the same (fast) university network, negligible effects of lag 
were experienced.

\systemname was used to transfer a randomly generated (via \texttt{/dev/ 
urandom}) 100KB binary file from one player to another. Network
traffic generated by the game was captured using \emph{Rawcap}~(a command-line
raw socket packet sniffer for Windows) with additional processing done using
{\tt tcpdump} on Linux.

We considered the impact of command rate (\ie how long AutoHotkey waits between
each command) and the impact of the maximum number of buildings selected ($k$) 
on the performance and security of \systemname. For this we varied the command 
delays from 100 ms/command to 1000 ms/command. In the same vein, the number of 
selected buildings is varied from 25 to 200. Additionally, to observe the impact 
of game-specific modifications, we evaluated the throughput of \systemname over 
the closed-source Aeons, with and without any game-specific modifications, in 
the same settings described above.

In order to compare the traffic characteristics of \systemname with
characteristics of the standard game, we gathered network traces of regular
0 A.D. two-player games. These were also collected in a similar setting --
i.e., with both players on the same university network and via direct connect.
Ten traces were collected (one per game played). Each of the recorded games
was between 20 and 60 minutes long.

\section{Security Evaluation}\label{sec:security}
We now perform an evaluation of \systemname against the network adversary
described in Section \ref{sec:threat}.

\subsection{Resilience to network traffic attacks}
\myparab{Passive analysis. }We first consider attackers with the
ability to perform IP and port filtering, deep-packet inspection, and
simple flow-level statistical analysis at line rate.

\textit{IP and port filtering: }Since \systemname actually uses an
off-the-shelf implementation of the game application, the IP address and 
ports used by \systemname are identical to that of the standard use of the
game. This means that an adversary that triggers blocking based on the 
destination IP (e.g., the game server) or port number, will be forced to
block all traffic to and from the game being used as the cover protocol.

In the event that the censor is willing to block \emph{all connections} to
dedicated game servers (often hosted by game publishers -- e.g., Electronic
Arts, Microsoft, Blizzard, etc.), clients may still utilize \systemname in 
direct-connect mode, forcing the censor into a game of
whack-a-mole with \systemname proxies hosted outside their jurisdiction.
Further, users may also easily migrate \systemname to another \rts game 
whose game servers are unblocked.

It is also worth noting that blocking game flows is not without costs to the
censor, specifically with respect to political good will and PR internationally.
For example, blocking all traffic for a given game, especially a popular game,
may upset citizens within their country and reflect poorly on Internet freedom
within the censoring country \cite{banned-1, banned-2, banned-3}.  

\textit{Deep-packet inspection:} When used with games that encrypt
their communications, \systemname is resistant to deep-packet
inspection, since the censor cannot decrypt the stream of moves being
made.

However, since \systemname works by issuing only generic commands (e.g., 
\texttt{MOVE} and \texttt{SET-RALLY-POINT}), it can easily be detected by 
DPI boxes if the game communicates commands in plain-text. Fortunately, most 
commercial \rts games perform command channel encryption (e.g., all of the 
Top 10 best-selling \rts games), making them resilient to such censors.

\begin{figure}[t]
%        \label{fig:ks-size}
        \centering
        \begin{subfigure}[b]{.48\textwidth}
                \includegraphics[trim=0cm 0cm 0cm
1cm,clip=true,width=\textwidth]{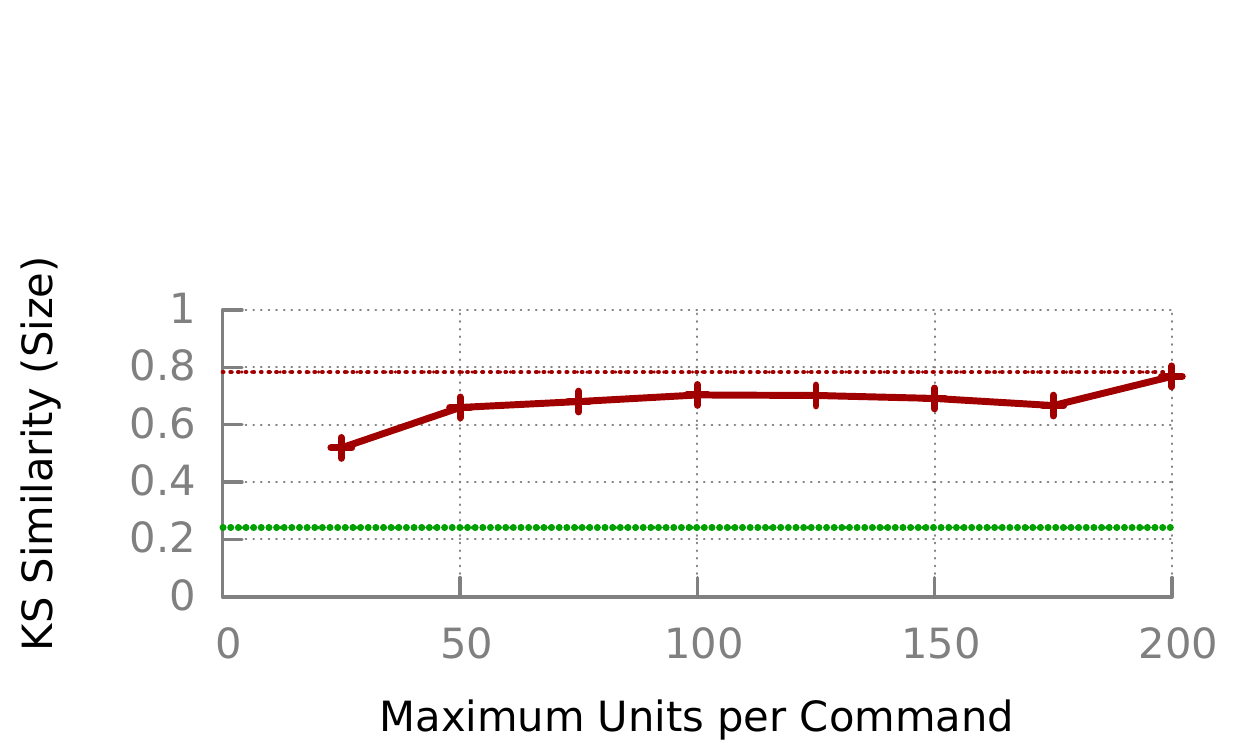}
               \label{fig:ks-size-units}
        \end{subfigure}%
	\vspace{-.6in}
        \begin{subfigure}[b]{.48\textwidth}
                \includegraphics[trim=0cm 0cm 0cm 1cm,
clip=true,width=\textwidth]{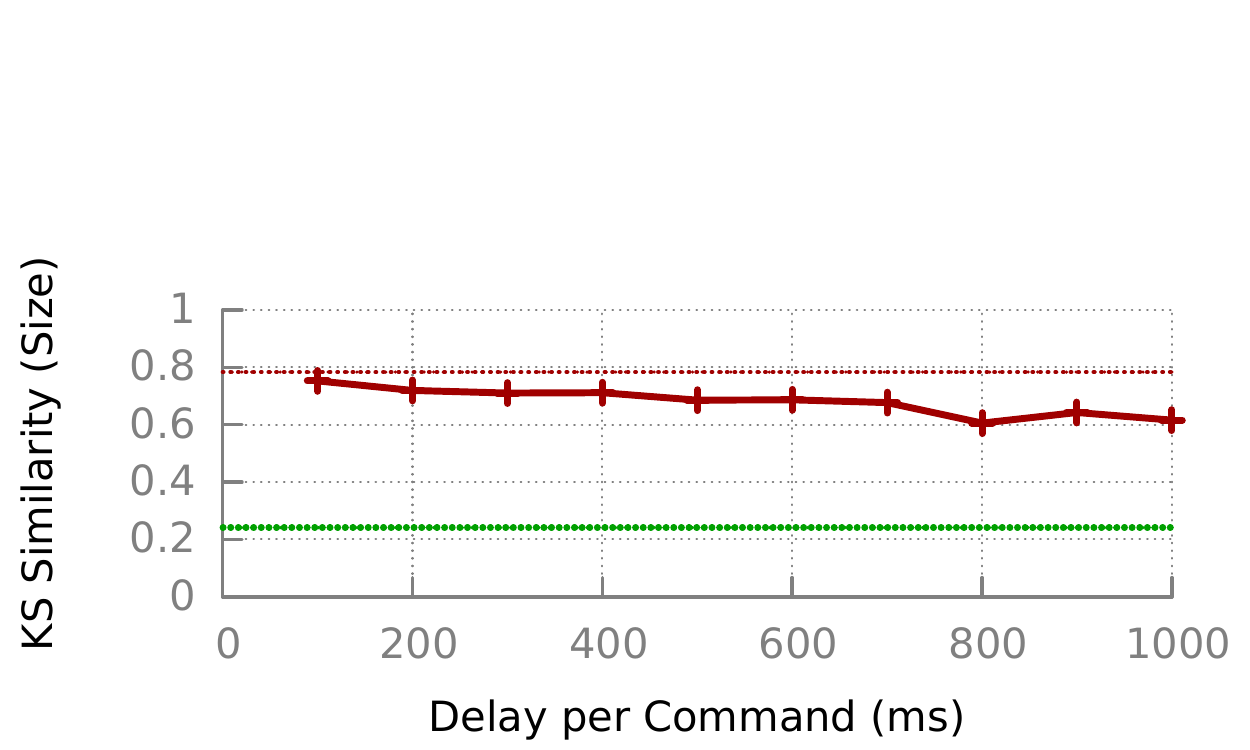}
	        \label{fig:ks-size-clicks}
        \end{subfigure}
	\vspace{-.5in}
        \begin{subfigure}[b]{.48\textwidth}
	\vspace{-1in}
                \includegraphics[trim=0cm 0cm 0cm
1cm,clip=true,width=\textwidth]{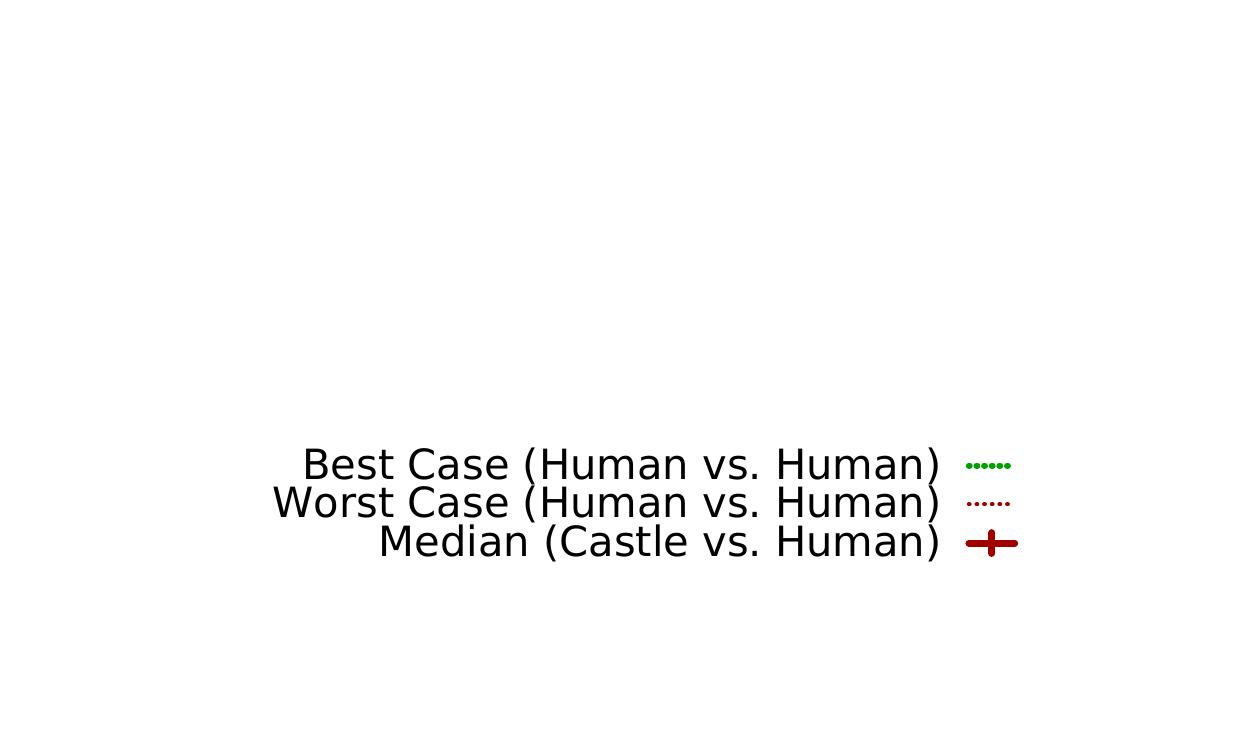}
        \end{subfigure}
	\vspace{-.2in}
        \caption{Kolmogorov-Smirnov (KS) statistic on the distributions of
packet sizes. The difference between \systemname and the
legitimate game flows is within the variance observed when comparing traffic
between legitimate game flows.}
        \label{fig:ks-size}
\end{figure}

\begin{figure}[t]
        \centering
        \begin{subfigure}[b]{.48\textwidth}
                \includegraphics[trim=0cm 0cm 0cm
1cm,clip=true,width=\textwidth]{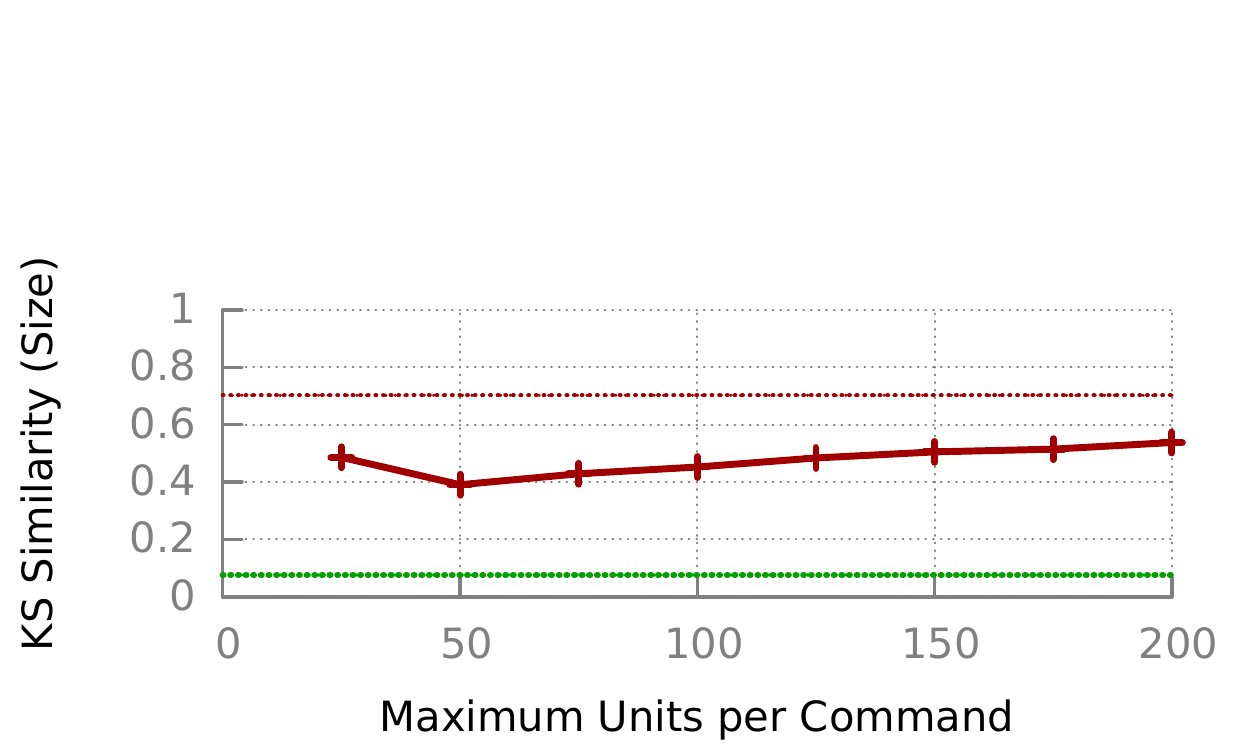}
                \label{fig:ks-time-units}
        \end{subfigure}
	\vspace{-.6in}
        \begin{subfigure}[b]{.48\textwidth}
	\vspace{-.6in}
                \includegraphics[trim=0cm 0cm 0cm
1cm,clip=true,width=\textwidth]{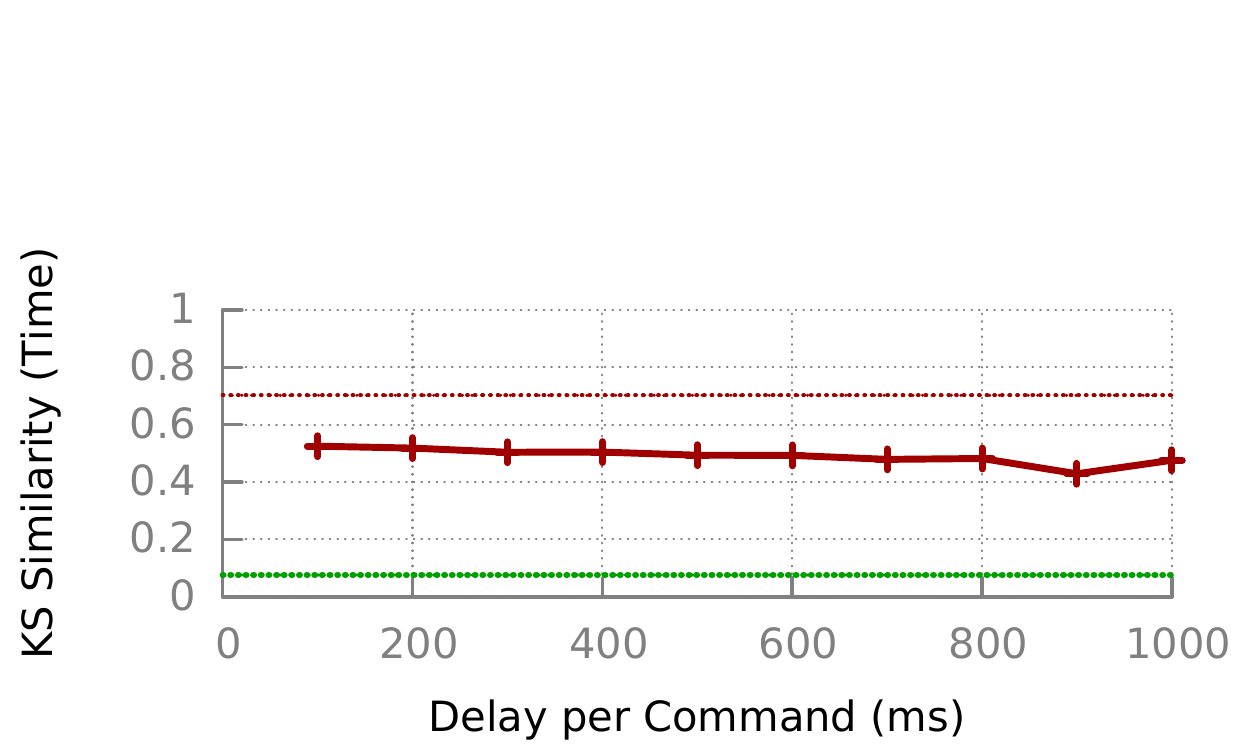}
                \label{fig:ks-time-clicks}
        \end{subfigure}
        \begin{subfigure}[b]{.48\textwidth}
	\vspace{-.4in}
                \includegraphics[trim=0cm 0cm 0cm
1cm,clip=true,width=\textwidth]{./ks-key.pdf}
        \end{subfigure}
	\vspace{-.7in}
       \caption{KS statistic on the distributions of
inter-packet times. The difference between \systemname and the
legitimate game flows is within the variance observed when comparing traffic
between legitimate game flows.}
        \label{fig:ks-time}
\end{figure}

\textit{Flow-level statistical analysis: }To quantify the resilience of 
\systemname against flow-level attacks, several statistical tests and 
classifiers were employed. For each experiment, the \systemname configuration
parameters that control the command rate and the maximum number of buildings 
selected in a command were varied between 0-1000 ms and 25-200 buildings, 
respectively.

First, the \emph{Kolmogorov-Smirnov} (KS) statistic was used to compare the 
similarity of human-game-generated traffic and \systemname-generated traffic. 
%In addition, the KS statistic was also used to understand the variation in 
%flow-level features of human generated traffic. The results are illustrated 
%in Figures \ref{fig:ks-size} and \ref{fig:ks-time}. 
Figure \ref{fig:ks-size} reflects the KS similarity statistic on the packet 
size distributions of human- and \systemname-generated games and Figure 
\ref{fig:ks-time} does the same for inter-packet times. %The heat-maps reflect 
%the density of (human, human) game pairs had the corresponding KS similarity
%score and the black crosses represent the KS similarity score between (human,
%\systemname) game pairs. The \emph{hot} regions on the plot indicate that there
%are a higher number of (human, human) game pairs with the corresponding KS score
%and the \emph{cold} regions indicate the opposite. 
We make two observations 
from these plots: (1) There is a high variation in the flow-level features of
legitimate (i.e., human-game-generated) traffic. We hypothesize that this is
because the traffic generated by the \rts game is strongly 
dependent on many parameters such as map and scenario type, strategies employed, and
number of players. (2) \systemname in many 
configurations, generates traffic that is well within this variance. We find that 
while restricting the maximum number of units per command to under 50 and the 
command rate to around 1 command/second, \systemname generates traffic that is 
as most similar to traffic generated by legitimate games.

Next, \systemname was evaluated against several website and traffic 
fingerprinting classifiers. The goal was to evaluate the accuracy of 
classifiers, built for flow-level analysis, in distinguishing between 
\systemname-generated and human-generated traffic. 

First, each network capture was split into (20) one minute long chunks. For each 
experiment, classifiers were given 20 chunks of \systemname-generated traffic 
at a specific configuration and 20 randomly selected human-game-generated 
chunks. Ten-fold cross validation was employed for splitting chunks into 
training and testing sets. 

Since, in our experiments, \systemname was used for the purpose of 
file transfer, all traffic generated by it was in a single direction. This
makes it trivially detectable by some fingerprinting classifiers which are 
heavily reliant on burst and direction features (e.g., k-NN\cite{knn}, the 
Panchenko classifier \cite{panchenko} and OSAD\cite{osad}). We note that 
in a real deployment this directionality would not be an issue as there would 
be requests/responses from both sides. 

\begin{figure}[t]
        \centering
	\begin{subfigure}[b]{.48\textwidth}
                \includegraphics[trim=0cm 0cm 0cm
1cm,clip=true,width=\textwidth]{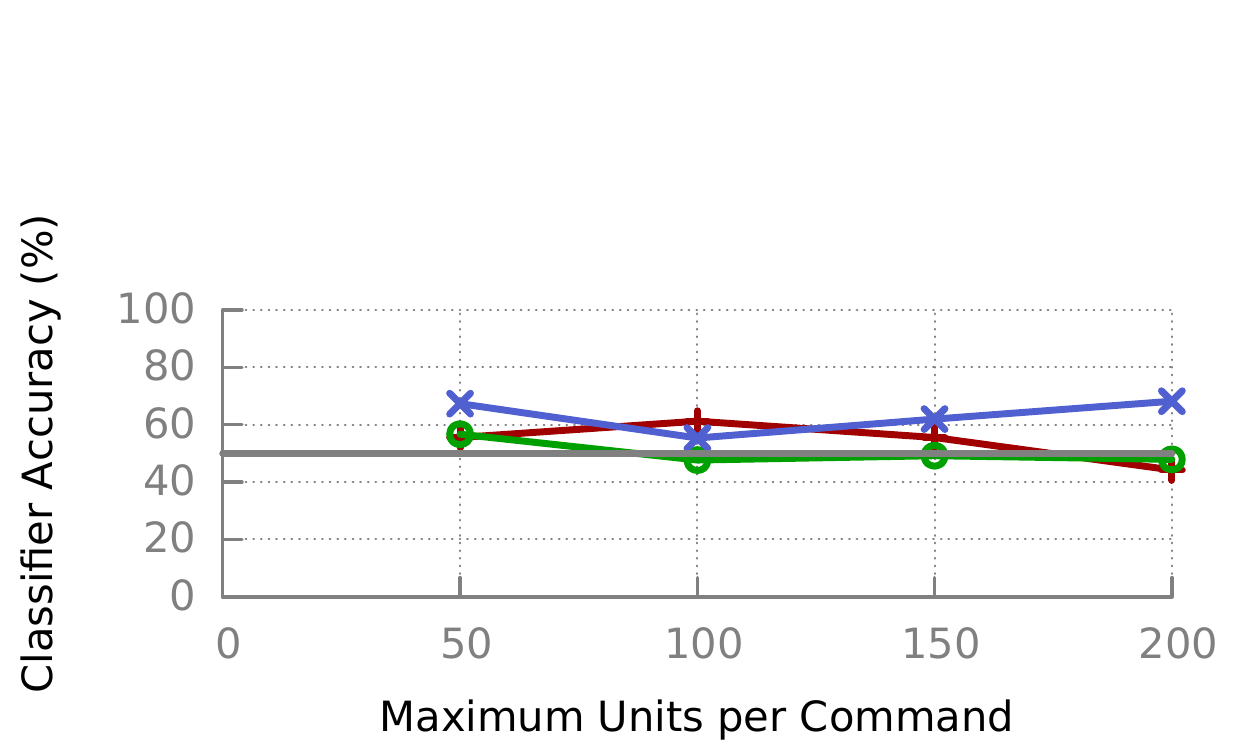}
		\label{fig:classifier-units}
        \end{subfigure}%
	\vspace{-.5in}
        \begin{subfigure}[b]{.48\textwidth}
                \includegraphics[trim=0cm 0cm 0cm
1cm,clip=true,width=\textwidth]{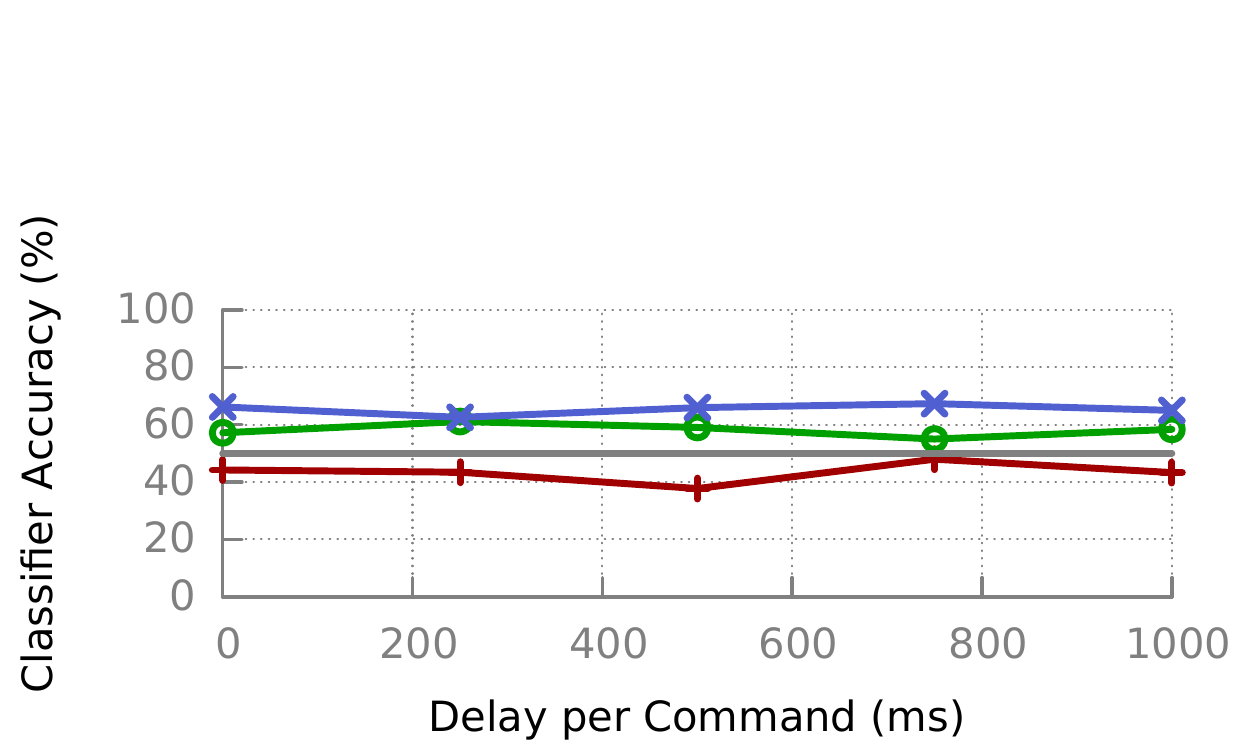}
                \label{fig:classifier-clicks}
        \end{subfigure}%
%        \label{fig:classifier-results}
\vspace{-1.05in}
        \begin{subfigure}[b]{.48\textwidth}
                \includegraphics[trim=0cm 0cm 0cm
1cm,clip=true,width=\textwidth]{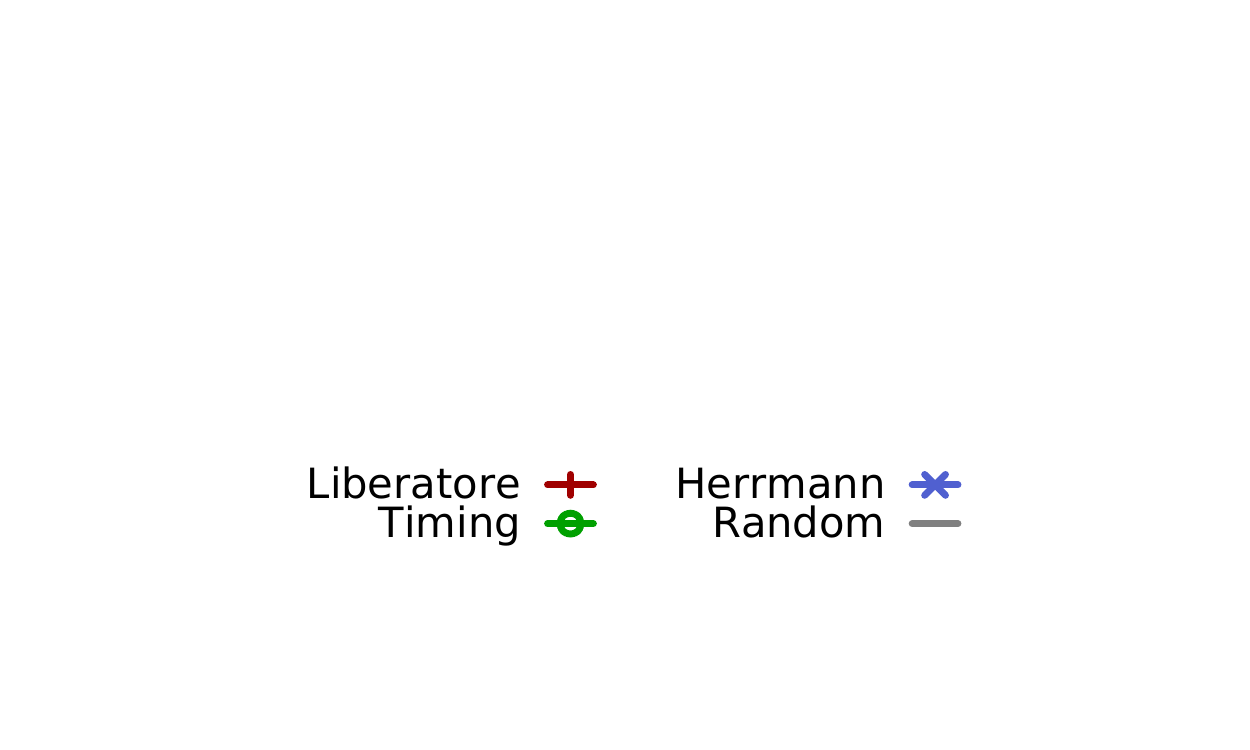}
        \end{subfigure}
	\vspace{-.65in}
        \caption{The performance of \systemname in various configurations 
against website fingerprinting classifiers.}
	\label{fig:classifier-results}
\end{figure}

Due to the directionality of traffic, website fingerprinting
classifiers that ignored directional information were used. These
included the Liberatore classifier \cite{liberatore}, the Herrmann
classifier \cite{Herrmann}, and an inter-packet timing classifier
\cite{Shmatikov}. All classifier implementations were obtained from
Wang's open-source classifier archive\cite{tao-wf-page}. The results
of these experiments are illustrated in Figure
\ref{fig:classifier-results}. In general, the results indicate that
\systemname performs very well against packet size and timing
classifiers, with only the Herrmann classifier achieving an accuracy
of over 60\% against multiple configurations of \systemname.  This is
not much better than random guessing.
%% Thes results suggest that, when used for bi-directional communication,
%% \systemname 
%% implemented as a web proxy will likely perform reasonably well even against 
%% classifiers considering burst information.

%\begin{figure}[t]%{.48\textwidth}
%	\centering
%	\includegraphics[trim=0cm 0cm 0cm 0cm,clip=true,width=.48\textwidth]
%	{./plots/mp-classifier-results.pdf}
%	\caption{Effect of \systemname Configurations on Classifier Accuracy}
%	\label{fig:classifier-results}
%\end{figure}

%%PGNOTE: please watch out for non-technical phrases
% "packet mangling" does not mean something specific in the technical sense.
\myparab{Active traffic manipulations.} In the face of active traffic manipulation attacks,
such as probing, packet injection, and  modification, \systemname 
implemented over most commercial games faces little threat. 

\textit{Packet injection. }If \systemname is implemented over a 
\rts game with an encrypted and authenticated command channel (e.g., any of the 
Top 10 best-sellers of all time), any packets injected by an unauthenticated source 
are dropped by the game-server. As a result, a probing adversary learns nothing 
about the \systemname games running on the server.

\textit{Packet modifications. }Most 
packet modification attacks are  prevented by the presence of encrypted 
and authenticated in-game channels. 
%% Even for authenticated clients, the 
%% constant in-game state verification prevents unexpected changes to the 
%% \systemname game state via modified packet contents and headers.
Additionally, since \systemname does not require any changes to the game or the
hosting server, such attacks will always elicit the same response from both,
legitimate game players and \systemname users.

\textit{Packet dropping and delaying. }Although most commercial \rts games make
use of UDP as a transport, the presence of reliability implemented in the 
application layer prevents any threats from adversaries that drop, or 
significantly delay packets in transmission. As a result, attacks (\eg~\cite{cover-your-acks})
that result in denial-of-service for \systemname users 
are not possible without also affecting legitimate game players.

\subsection{Resilience to application layer attacks}

Highly motivated censors may perform actions outside the realm of standard 
network traffic analysis and manipulation. We consider censors that may interact 
with the game server using custom game clients in order to reveal the identities
of \systemname users. Specifically, censors may connect to game server lobbies 
to identify \systemname games and join these games to learn the IPs of 
participating  clients. For these cases, \systemname provides several defenses 
based on features available in the cover game.

If the cover game supports the use of password-protected multi-player games, 
\systemname proxies may host games using high-entropy passwords distributed 
using, for example, a BridgeDB-like mechanism \cite{TorDB}. Therefore censors 
without knowledge of the password are unable to join \systemname games and 
learn the IP addresses of \systemname clients.

If the cover does not support the use of password-protected games, a \systemname
proxy may incorporate either (or, both) of the following defenses: (1) The proxy
may use standard game maps rather than custom-made \systemname game maps. This
allows \systemname instances to blend in with legitimate game instances, making
it harder for the censor to identify which games to join. However, this comes at
the cost of lower throughput since there are typically fewer units in standard
game maps. (2) The proxy may still use a BridgeDB-like mechanism for password 
distribution and require that any \systemname client makes the moves 
corresponding to the supplied password in order to receive proxying services. 
In the event that a client does not supply this password within some period 
of time, the \systemname proxy may continue playing the game using a standard 
AI. Therefore, even a censor that is able to enter games is unable to 
distinguish between \systemname games and legitimate player games.

\subsection{Avoiding covert channel pitfalls}

Geddes \etal highlight three key mismatches between covert channels
and cover traffic which make these look-like-something circumvention tools
detectable to external observers~\cite{cover-your-acks}. Here we discuss how
\systemname avoids each of these three mismatches.

\myparab{The architecture mismatch.} Games provide agility in terms of
architecture that few other channels provide. They often operate in
client-server mode on publisher-hosted game servers and in peer-to-peer mode in
direct-connect multi-player games. Our proxying approach can operate in
whichever mode is the dominant, and in the
presence of blocking can even shift (\eg from client-server to peer-to-peer).

 \myparab{The channel mismatch.} While game data is
typically communicated over a UDP channel, it is %unlike most other UDP channels.
%Unlike channels such as VoIP, game traffic is 
not resilient to packet loss like other UDP-based channels (\eg VoIP), thus
clients come with the ability to handle packet losses and retransmissions.
Further, they also guarantee in-order delivery and processing of sent data. This
makes it especially useful as a covert channel for proxied TCP connections which
require reliable transmission. Therefore, attacks that allow the censor to drop
traffic to levels which are tolerable to legitimate players (but intolerable to
\systemname users) are not possible.
%Further, active modification of communicated 
%content, by an adversary, is easily detected and ignored by the game state 
%machine and cheat detection modules. 

\myparab{The content mismatch. }Content mismatches arise when the content being
embedded in the covert channel changes the flow-level features of the channel.
Since the flow-level features of \rts games are strongly dependent on many
parameters (identified above), they are highly variable. We have shown that
\systemname, under every configuration, generates traffic that is well within
this variance.

\section{Performance Evaluation}\label{sec:performance}

Without any game-specific modifications, \systemname offers performance amenable
to transfer of textual data (\eg tweets, e-mail, news articles).\footnote{The
success of the voices feeds\cite{the-voices}
%http://johnscottrailton.com/the-voices-feeds/
during the Arab Spring shows that in some situations textual data is enough to
get information out.} 

Since each \rts game has a limit on the number of objects 
that can be selected for a single command, the data rate obtained by \systemname
is game dependent. For example, 0 A.D. allows the selection of up to 200 units
for a single command, giving us an average of
$\frac{\sum_{i=1}^{200}\log_2\binom{1600}{i}}{200\times 8} \approx 65$ bytes per
command. On the other hand, Aeons has no limits on the number of units that
may be selected for a single command but allows only $\leq$ 435 objects 
to be placed within a single screen -- giving us an average of $\approx 39$
bytes 
per command.

Throughput is also dependent on the time required by the desktop automation tool
to perform the actions required to issue a command (i.e., click 
%click each and every unit (ie., click all units) or click a single unit?
each unit to be
selected and click the target coordinate). We found that on average, issuing a
single command required between 300-350 ms. With no delays between the issue of 
each command, this allows $\approx$ 3 commands/second.

%%\rish{Realized this was garbage. Taking it out.}
%Since \rts games often batch user issued commands into
%fixed time slots \Pnote{does this mean 1 set of building selections + rally
%point per interval, or multiple selections and rally points batched together?},
%the data rate provided by \systemname is game dependent. For example, 0 A.D.
%batches user issued commands into time periods of between 200-400 ms --
%allowing
%at most 2-3 \systemname generated commands to be issued per second (each
%command
%providing an average of 65 Bytes with our current encoding). For games with
%smaller batch times (e.g., 2 of the top 10 best selling \rts games are known to
%batch commands into 100-200 ms intervals), the data rate increases linearly.
%\Pnote{ can we add those lines to the plot? E.g., "Game with 100 ms batching"
%(or better legend title)}
%

\begin{figure}[t]
        \centering
        \begin{subfigure}[b]{.48\textwidth}
                \includegraphics[trim=0cm 0cm 0cm 1.8cm, clip=true,
width=\textwidth]{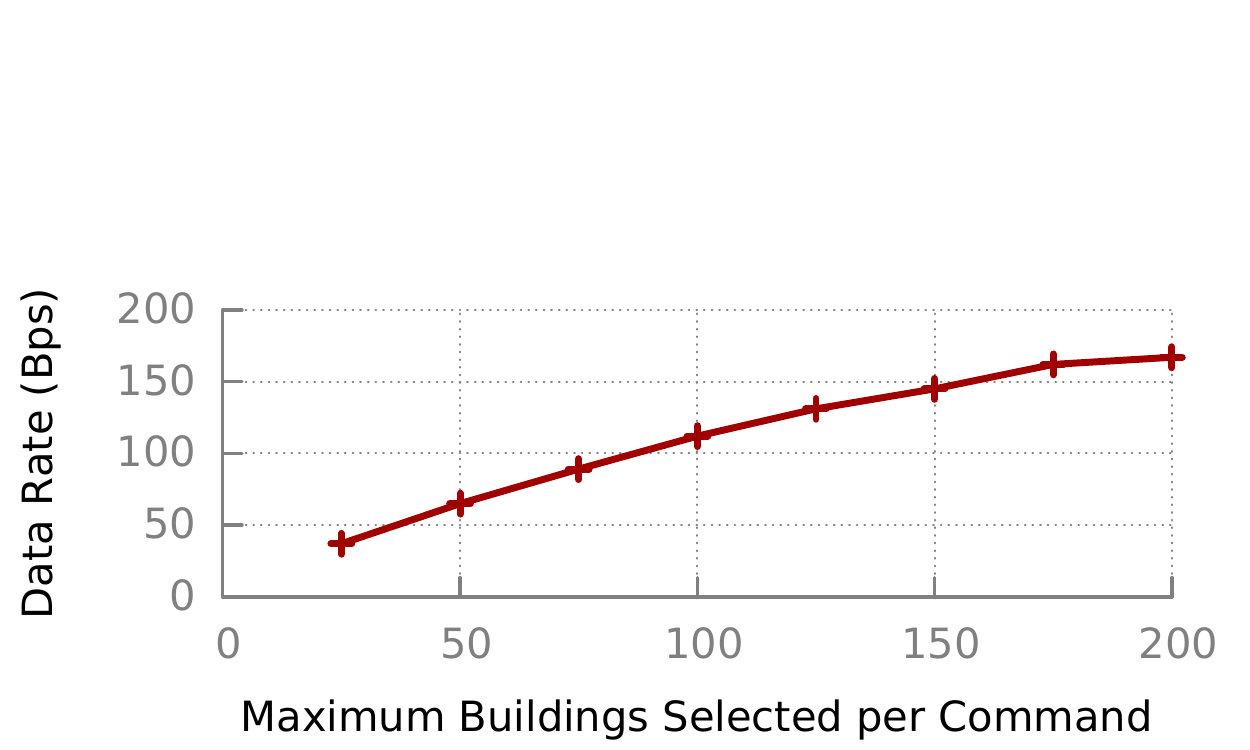}
                \caption{Effect of maximum number of buildings selected per
command (at $\approx$ 100 ms delays/command)}
                \label{fig:xput-units}
        \end{subfigure}%

        \begin{subfigure}[b]{.48\textwidth}
                \includegraphics[trim=0cm 0cm 0cm 1.8cm, clip=true,
width=\textwidth]{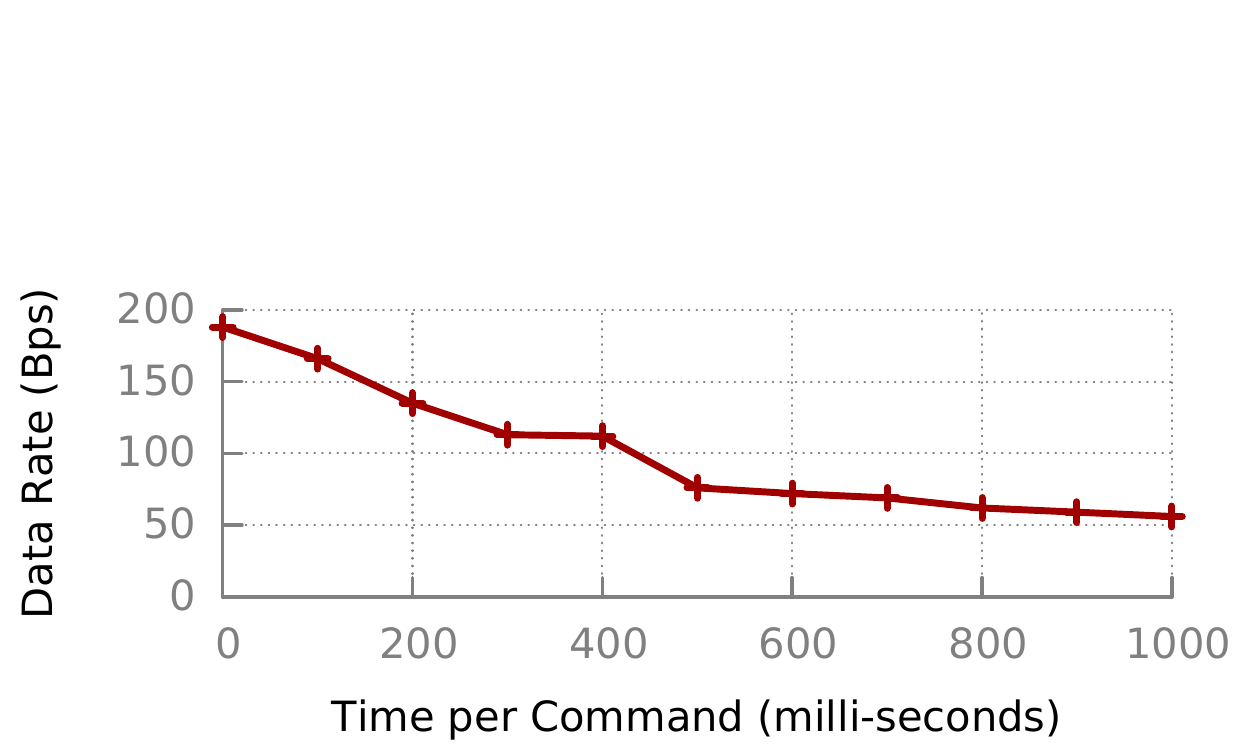}
                \caption{Effect of time delays per command (at $\leq 200$
buildings
selected/command)}
                \label{fig:xput-clicks}
        \end{subfigure}
        \caption{Throughput of \systemname implemented over 0 A.D. under various
configurations}
	\label{fig:xput}
\end{figure}

In Figure \ref{fig:xput}, we see the effect of \systemname's parameters on it's
performance when implemented over 0 A.D. Specifically, Figure
\ref{fig:xput-units} shows the effect of increasing the maximum number of
buildings selected in a single command and Figure \ref{fig:xput-clicks}
demonstrates the effect of decreasing the command rate. At a configuration where
\systemname may select up to 200 buildings in a single command and issues
commands 
with no delays in between, \systemname implemented over 0 A.D. is able to
provide 
a data rate of $\approx$ 190 Bytes per second -- requiring about 52 seconds for 
the transfer of a short 10KB text news article.

\subsection{Game-specific enhancements for \systemname}\label{subsec:sc2}
In this section, we show that the performance of \systemname can be improved
significantly through simple game-specific tweaks. To be able to observe the
impact of these game-specific modifications, Aeons was used as the channel
for \emph{vanilla} \systemname and \systemname with Aeons-specific
modifications. The game-specific modifications were introduced and implemented
for \systemname in just under 
%nit: this is actually the correct approach (spell out all numbers smaller than
%10). please apply to rest of draft :)
three hours by an undergraduate researcher.

The low throughput of \systemname over Aeons
was because Aeons had larger units than 0 A.D., thereby allowing players to
place only
435 units within a single screen (as opposed to 1,600 for 0 A.D.). As a result,
the throughput of vanilla \systemname was only $\approx 38$ bytes/command (i.e.,
$\leq $ 130 bytes/second) at best -- i.e., with the maximum command rate of
AutoHotkey and 
selection of up to 435 units/command. 

A quick investigation into the Aeons replay mode and save-game 
files revealed that even the selection of a single unit was communicated over
the network 
and logged by other players. We exploit this fact by creating a set of $2^m$
units (256 in our case) and mapping
each unit to an $m-bit$ value (\ie a byte). We then sequentially transfer the
data byte-by-byte via selecting the unit
corresponding to the byte to be encoded. 

%This allowed a minor change to \systemname's encoding 
%that resulted in significantly better throughput. Specifically, following this 
%observation, the undergraduate introduced the following 
%changes to \systemname's encoding: A $n = 2^m$ ($m \leq 8$) unit map was
%created in 
%which each unit was used to represent $m$ bit values. In order to encode a
%particular $m$ bit chunk of covert data, the corresponding unit was selected.
%Therefore, every selection by \systemname resulted in the transfer of $m$ bits 
%of covert data. This process was repeated until all the covert data was encoded 
%as \emph{select unit} commands. 

\begin{figure}[t]
\centering
\includegraphics[trim=0cm 0cm 0cm
2cm,clip=true,width=.48\textwidth]{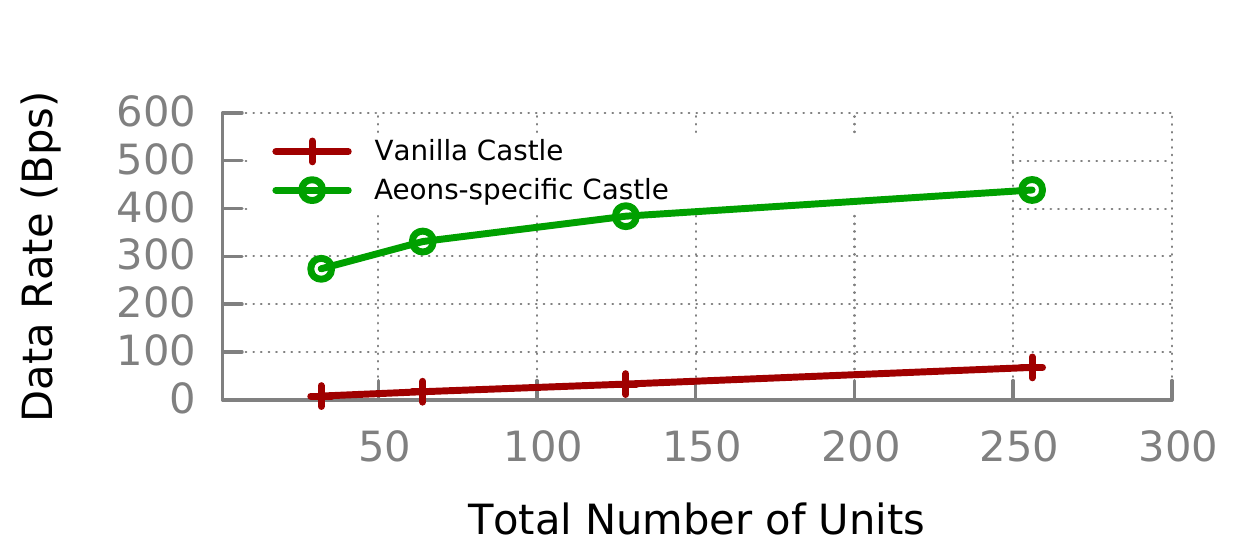}
                \caption{Throughput of \systemname (with and without
enhancements) implemented over closed-source Aeons}
                \label{fig:starcraft-xput}
\end{figure}

This encoding allowed AutoHotkey to issue
commands at a significantly faster rate than before (a command was now just a
single mouse click, as opposed to up to 435 key presses and clicks). At
AutoHotkey's 
fastest mouse click rate and $m=8$, this encoding achieved a throughput of up to 
3KByte/second. However, in order to more closely mimic the command rate and
traffic 
generated by a human player, we add a delay of between 2 and 3 $ms$ per command.
%\rish{We
%don't have real traffic to compare to :(}
In Figure \ref{fig:starcraft-xput}, we show the % dramatic - poetry.
effect of this game-specific modification on the throughput of \systemname. From
the same figure, we can also observe the effect of varying the total number of
units with
vanilla \systemname and the Aeons-specific version of \systemname. We see
that 
increasing $n$ results in a linearly increasing throughput for vanilla
\systemname, 
and a logarithmically increasing throughput for Aeons-specific \systemname.
However, because the cross-over point of these functions is higher than the game
allows, 
Aeons-specific \systemname always achieves better throughput for Aeons.

\section{Discussion}\label{sec:discussion}

In this section, we discuss \systemname in the context of its ability to 
provide deniability to users of the system and provide extensibility that can 
tip the scales in favor of circumvention developers. We also discuss various
methods to improve the throughput of \systemname and compare the design 
methodology of \systemname with the most similar related work -- Rook and
FreeWave.

\myparab{Deniability and ease of distribution. } One of the advantages of
\systemname is that the covert channel is largely implemented with off-the-shelf
software components with only a few hundred lines of code dedicated to encoding 
data and desktop automation scripting. Desktop automation tools, are already 
commonly used by gamers and the game, game-specific mods, etc., are generally 
widespread enough to warrant little suspicion from censors (\eg Aeons is 
installed by millions of users). 

\systemname's small core size also makes it easy to distribute through hard to 
block methods -- e.g., via the text body in emails and even through instant 
messaging services.

\myparab{Extensibility of \systemname. }\systemname's strength comes from the
ease with which it can be ported to new games. As an example, it took a bright
undergrad less than 6 hours to complete a basic port of \systemname over a very
popular closed-source \rts game. Due to the availability of game-specific hacks
and reverse engineering guides in popular gaming forums~\cite{hacking-forum, 
hacking-forum-1, hacking-forum-2}, completing game-specific enhancements in order 
to improve the data rate of \systemname, as described in Section~\ref{subsec:sc2}, 
required only an additional 3 hours. 

Although individual game titles do not present a high collateral damage, in the
event that they are blocked, \systemname presents a simple way to convert each
of them into an ephemeral and effective covert channel, with little development
overhead. This ability, along with the fact that every newly released title is a
potential covert channel, makes \systemname particularly useful in the arms-race
that censors and developers are currently engaged in. In particular, it is the
first censorship circumvention tool to provide an asymmetry in favor of the
developer (i.e., creating a new channel is significantly less expensive than
detecting the channel).

\myparab{Improving \systemname throughput. } In addition to making game-specific
modifications, \systemname presents many opportunities to increase throughput of
the system. 

\begin{packeditemize} \item \emph{Parallel requests: }Since most modern \rts
games allow up to eight players to participate in a single multi-player game, it
is possible for one censored user to decode content responses from up to seven
proxies in parallel -- achieving up to $7x$ downstream throughput. This is
particularly useful in the context of web data, where requests are easy to
parallelize. 

%\item \emph{Game Specific Enhancements: }Many \rts games offer features that
%are not universal. For instance, many games provide \emph{trigger} controls to
%map designers -- i.e., a feature that allows map designers to specify responses
%to player actions (if a player performs action $x$, action $y$ happens to unit
%$z$). Such features allow \systemname to encode significantly more data than
%currently possible -- e.g., \systemname could use a hierarchical encoding
%structure if camera motion actions are permitted in trigger systems. Other
%games provide significantly more comprehensive replay information and include
%preserving the order of clicks performed by opponents. This allows castle to
%achieve significantly more Bytes per command ($O(\log_2P(n, k))$) than it
%currently does ($O(\log_2C(n, k))$).

\item \emph{Content compression. }\systemname proxies may improve performance by
compressing requested content before encoding. In the context of web data, the
proxies may also pre-render and compress content before sending them to the
receiver (\eg as was done by the Opera mobile browser~\cite{opera}).  \item {\em
Trade-off throughput and detectability.} Depending on the level of surveillance
in a given region \systemname may expose an option to allow users to trade off
throughput \vs detectability of the system (\eg by increasing the rate of clicks
in the automation tool). 
\end{packeditemize}

\myparab{Comparison of design methodology. }In terms of design
methodology, Rook and FreeWave are most similar to \systemname. Like
\systemname, Rook also uses games as the cover protocol, although the
goals of Rook and \systemname are quite different.  The primary goal
of \systemname is adaptability -- \systemname is loosely coupled to
the underlying game, enabling developers to quickly adapt \systemname
to many games.  Rook is focused on stegonographic security; even if
the adversary is able to join the same gaming session as two Rook
users, the attacker will still not be able to determine whether the
other players are using Rook to surreptitiously transmit data.  As a
result, Rook aims for security against a very powerful adversary, but
has over 100x lower bandwidth than \systemname.

%% General advice: don't trash other people's research
%%
%% . While Rook also has the property of being 
%% extensible to other games, it requires an understanding of how games encode 
%% data in network packets, and knowledge of which fields are mutable. Such 
%% information is hard to find for many closed-source games, which 
%% significantly impacts its extensibility. On the other hand, \systemname only 
%% requires the availability of replay file decoders. These are much more widely 
%% available since they are often used and are of interest to many casual gamers.

FreeWave and \systemname are similar in that they both work above the
application layer, mimicking user input to the application, rather than the 
application itself (FreeWave uses a modem to encode data into audio 
transmissions over VoIP clients). 
%% This is not supported by any evidence, and I suspect is not even true. -- rob
%%
%% In this context, \systemname benefits from 
%% the fact that in-game actions by players provide significantly higher entropy
%% than human voice, therefore providing a covert channel that is harder to
%% detect. 
Similar to \systemname, FreeWave also allows extensibility to other
similar applications. However, there are significantly more \rts games available
for use as cover protocols than there are VoIP clients.

To the best of our knowledge, \systemname is the only covert channel
that (1) appears to satisfy all the covert channel design principles
laid out by Geddes, et al\cite{cover-your-acks}, (2) provides
extensibility to hundreds of applications (games), potentially with
only a few hours effort for each, and (3) is evaluated for security at
the application and network layer.

%\myparab{Raising the Bar for Censors: }\rish{Not exactly sure how to define the
%cost to the censor for doing the analysis we did. Leaving this for now, I'll
%come back to it later.}

%Things to talk about here ...
%
%\myparab{Increasing throughput.} multiple clients. compress before coding.
%different types of game (e.g., realtime, racing).
%
%\myparab{Extensibility of the system.} Games are likely not high collateral
%damage to block. This system easily can adapt to new video games make it
%amenable to arms racing + winning. Also how we extend the data receiving
%process. reve rse engineering. datapoint on games we've got working + link to
%forums where people already do this.
%
%\myparab{Raising the bar for censors.} Some back of the envelope calculations
%about how hard the analysis in evaluation would be at line rate in an ISP.
%Also, how hard would it be for an ISP to identify this sort of thing.
%

%\myparab{Another advantage but I don't know where to put it.} Another advantage
%here is that aside from the translation layer the user isn't installing much in
%the way of special circumvention software. Just the game and the desktop
%automation which are general.

\section{Conclusions}\label{sec:conclusions}

In this paper we have presented \systemname, a general approach for creating covert channels using \rts 
games as a cover for covert communications. We demonstrate our approach by prototyping on two different games 
with minimal additional development overhead and show its resilience to a network adversary. 
%In this paper, we have motivated the use of video games as a covert channel. 
%Through \systemname and the \rts genre, 
We argue that the %architectural 
%agility, security and reliability, 
popularity, availability, and
generic functionalities of modern games make them an effective circumvention 
tool in the arms-race against censors. Specifically, our results show that 
\systemname is:

\begin{packeditemize}
\item \emph{Secure: } \systemname is resistant to attacks such as IP/port
filtering and deep-packet inspection since it actually executes the game
application. More complicated and expensive attacks such as traffic analysis
attacks are avoided due to the high variability of standard game flows.
\item \emph{Usable: } Even without any game-specific modifications, \systemname
is able to provide throughput sufficient for transfer of textual data.
Additional enhancements make it suitable for use as a web proxy.
\item \emph{Extensible: }Incorporating new closed-source games as covert 
channels for \systemname requires only a few hours of developer time -- 
including the addition of title-specific enhancements for increased throughput.
\end{packeditemize}

The results presented in this work motivates two independent future research
directions. First, extending our work to different classes of games which may enable higher throughput rates 
(\eg racing games, first person shooters). Second, integrating the \systemname
approach into platforms to make it usable to users \eg via a Web browser plugin
or integration with the suite of Tor Pluggable Transports~\cite{tor-pt}. 

\iffalse %a little unbaked.
First, an investigation to find other applications (with higher
collateral) that lend themselves as naturally as video games and the \rts genre, 
to censorship circumvention.
Second, more fundamentally, a game-theoretic investigation to understand if numerous,
low-effort, possibly ephemeral covert channels (such as those provided by
\systemname) present more of a challenge and expense to censors than the few 
existing covert channels.% -- i.e., is short-lived diversity in the space and time dimension more
%valuable than longer diversity in the space dimension.
\fi

\myparab{Code and data release: }To ensure reproducibility of our results and
ease comparative evaluation, our implementation of \systemname is available at 
\url{https://github.com/bridgar/Castle-Covert-Channel}.

\myparab{Video demonstration: }A video demonstration of \systemname implemented
over 0 A.D. is available at \url{https://www.youtube.com/watch?v=zpjZJuvMhDE}.

\section*{Acknowlegements}
This material is based in part upon work supported by the National Science 
Foundation under Grant Number  CNS  -  1350720.  Any  opinions,  findings,  
and  conclusions or recommendations expressed in this material are those of 
the author(s) and do not necessarily reflect the views of the National 
Science Foundation.

{\small
\bibliographystyle{unsrt}
\bibliography{bibliography} 
}

\end{document}